\documentclass[12pt,preprint]{aastex}

\usepackage{epsfig}
\begin{document}

\title{Early Afterglows of Gamma-Ray Bursts in a Stratified Medium with a
Power-Law Density Distribution}
\author{Shuang-Xi Yi$^{1,2}$, Xue-Feng Wu$^{3,4,5}$, and Zi-Gao Dai$^{1,2}$}
\affil{$^{1}$School of Astronomy and Space Science, Nanjing University, Nanjing 210093, China; dzg@nju.edu.cn \\
$^{2}$Key Laboratory of Modern Astronomy and Astrophysics (Nanjing
University), Ministry of Education, China \\
$^{3}$Purple Mountain Observatory, Chinese Academy of Sciences,
Nanjing 210008, China\\
$^{4}$Chinese Center for Antarctic Astronomy, Chinese Academy of
Sciences, Nanjing, 210008, China\\
$^{5}$Joint Center for Particle Nuclear Physics and Cosmology of
Purple Mountain Observatory-Nanjing University, Chinese Academy of
Sciences, Nanjing 210008, China}

\begin{abstract}
A long-duration gamma-ray burst (GRB) has been widely thought to
arise from the collapse of a massive star, and it has been suggested
that its ambient medium is a homogenous interstellar medium (ISM) or a
stellar wind. There are two shocks when an ultra-relativistic
fireball that has been ejected during the prompt gamma-ray emission phase sweeps
up the circumburst medium: a reverse shock that propagates into the
fireball, and a forward shock that propagates into the ambient
medium. In this paper, we investigate the temporal evolution of the
dynamics and emission of these two shocks in an environment with a
general density distribution of $n\propto R^{-k}$ (where $R$ is the
radius) by considering thick-shell and thin-shell cases. A GRB
afterglow with one smooth onset peak at early times is understood to
result from such external shocks. Thus, we can determine the medium
density distribution by fitting the onset peak appearing in the
light curve of an early optical afterglow. We apply our model
to 19 GRBs, and find that their $k$ values are in the range of 0.4 - 1.4,
with a typical value of $k\sim1$, implying that this environment is
neither a homogenous interstellar medium with $k=0$ nor a typical
stellar wind with $k=2$. This shows that the progenitors of these
GRBs might have undergone a new mass-loss evolution.
\end{abstract}

\section{Introduction}

Since their first discovery in 1997, gamma-ray burst (GRB) afterglows
have been well understood (Wijers et al. 1997; Piran 1999; van
Paradijs et al. 2000; M\'esz\'aros 2002; Zhang \& M\'esz\'aros 2004),
and are usually explained as being due to the interaction of an
ultra-relativistic fireball with its surrounding medium. During such
an interaction, there are two shocks when a relativistic fireball
sweeps up the ambient medium: a forward shock (FS) that propagates
into the circumburst medium, and a reverse shock (RS) that propagates
into the fireball ejecta. The observed afterglow arises from the
synchrotron emission of swept-up electrons accelerated by the FS and
RS. GRBs can be classified into two types: short-duration
hard-spectrum GRBs, which may originate from the mergers of two
compact stars, and long-duration soft-spectrum GRBs, which may come
from the core collapse of massive stars. The circumburst medium
surrounding these two types of GRBs may be different, due to
their different origins. By assuming that GRB afterglows are produced
by the fireball interacting with the circumburst medium, we can
use GRB afterglows to probe their environments. In this paper, we
assume a circumburst medium with a general density distribution of
$n = AR^{-k}$. Such a circumburst medium is a homogeneous
interstellar medium (ISM) when $k=0$, and a typical stellar wind
environment for $k=2$. Much work has been done in terms of
theoretical afterglow lightcurves for the case of an ISM environment
($k=0$) (Sari et al. 1998; Kobayashi 2000; Panaitescu \&
Kumar 2004) and for the case of a typical stellar wind environment
($k=2$) (Dai \& Lu 1998a; M\'esz\'aros et al. 1998; Panaitescu \&
Kumar 2000, 2004; Chevalier \& Li 2000; Wu et al. 2003, 2004;
Kobayashi \& Zhang 2003; Zou et al. 2005).

Many early optical afterglows  have been detected in
the {\em Swift} era. The observations could provide important clues
about the properties of the ambient medium of GRBs. Li et al. (2012)
extensively searched for optical lightcurves from the literature,
and found that optical afterglows have different radiation
components. These emission components may have distinct physical
origins. In this paper, we consider smooth onset peaks in early
optical afterglow lightcurves. The onset of an afterglow
is assumed to be synchronous with the moment when the fireball is
decelerated by the surrounding medium. Liang et al. (2010) found 20
optical lightcurves with such smooth onset features. We probe the type of GRB
ambient medium with the rising and decaying slopes of the
onset peak in the optical lightcurve. We study the emission of reverse-forward
shocks both in the thick- and thin-shell cases for an environment
with a general density distribution of $n= A{R^{ - k}}$. We apply
our model to 19 GRBs as a case study and find a typical value of $k\sim1$ (see Fig. 5).
In \S 2 we discuss the hydrodynamic evolution of a fireball
in both thick-shell and thin-shell cases, and consider
reverse-forward shocks in each case. Theoretical lightcurves of
reverse-forward shock emission are derived in \S 3. We investigate
19 optical afterglow onset peaks in detail in \S 4. Discussion and
conclusions are presented in Sections 5 and 6, respectively. A
concordance cosmology with $H_0 = 71$ km s$^{-1}$
Mpc$^{-1}$, $\Omega_M=0.30$, and $\Omega_{\Lambda}=0.70$ is adopted.
$Q_n$ denotes $Q/10^n$ in cgs units throughout the paper.

\section{Hydrodynamics of a Relativistic Shell Interacting with Its Ambient Medium}

For a relativistic shell decelerating in its circumburst medium, two
shocks will develop: an reverse shock that propagates into the shell,
and an forward shock that propagates into the ambient medium. We
assume that the shell and two shocks are spherical and the shocked fluid
is uniform in the downstream. The shell is characterized by an
initial kinetic energy $E$, initial Lorentz factor $\eta$, and a
width $\Delta$ in the lab frame attached to the explosion center.
Physical primed quantities are defined in the comoving frame. The
co-moving number density of the shell is then ${n'_4} = E/(4\pi
{m_p}{c^2}{R^2}\Delta {\eta ^2})$, where $R$ is the radius of the
shell. The number density of the ambient stratified medium is
assumed to have the following general distribution, ${n_1} = A{R^{ -
k}}=n_0(R/R_0)^{-k}$. We fix $n_0=1$ cm$^{-3}$ and let $R_0$ be
variable. In this paper, we focus on the hydrodynamic evolution and
emission of the reverse-forward shocks in arbitrary stratified
ambient media with $0\leq k <3$. For $k \geq 3$, the
energy-conservation shock solution cannot be applied, and the
solution is limited between the shock front and the sonic point in
the downstream of the shock (Sari 2006). As in the literature, we
divide the two-shock system into 4 regions (Sari \& Piran 1995): (1)
the unshocked ambient medium ($n_1$, $e_1$, $p_1$, $\gamma_1$), (2)
the shocked ambient medium ($n'_2$, $e'_2$, $p'_2$, $\gamma_2$), (3)
the shocked shell ($n'_3$, $e'_3$, $p'_3$, $\gamma_3$), and (4) the
unshocked shell ($n'_4$, $\gamma_4=\eta$), where $n$ is the number
density, $e$ is the internal energy density, $p$ is the pressure,
and $\gamma$ is the bulk Lorentz factor. In the lab frame, the
ambient medium is assumed to be static, i.e., $\gamma_1=1$ (the
speed of the ambient medium  can be neglected in our problem). The
ambient medium and relativistic shell are assumed to be cold, i.e.,
the internal energy $e$ and pressure $p$ are negligible compared to
the rest-mass energy density $\rho c^2$. The shocked ambient medium
(region 2) and shocked shell (region 3) are assumed to have a
relativistic equation of state, i.e., $p'=e'/3$. The jump conditions
for the shocks are: $e'_2=(\gamma_2-1)n'_2 m_p c^2$,
$n'_2=(4\gamma_2+3)n_1$ for the forward shock, and
$e'_3=(\gamma_{34}-1)n'_3 m_p c^2$, $n'_3=(4\gamma_{34}+3)n'_4$ for
the reverse shock. The Lorentz factor of the reverse shock,
$\gamma_{34}$, can be approximated as
$\gamma_{34}=(\gamma_3/\gamma_4+\gamma_4/\gamma_3)/2$, as long as
$\gamma_3 \gg 1$ and $\gamma_4 \gg 1$. The equilibrium of pressures
and the equality of velocities along the contact discontinuity lead to
$p'_2=p'_3$ and $\gamma_2=\gamma_3$, respectively. To solve the
problem and using the initial conditions, we adopt the ratio of the
number density of the relativistic shell ${n'_4}$ to the number
density of the ambient medium ${n_1}$ defined in Sari \& Piran (1995),
i.e.,
\begin{equation}
f  = \frac{{{l^{3 - k}}}}{{(3 - k){R^{2 -
k}}\Delta {\eta ^2}}},
\end{equation}
where the Sedov length $l$ is defined when the rest-mass energy of
the swept ambient medium, ${M_{\rm sw}}{c^2}$, equals the initial
energy $E$ of the relativistic shell,
\begin{equation}
\quad l = \left[ {\frac{{(3 - k)E}}{{4\pi A{m_p}{c^2}}}}
\right]^{1/(3 - k)}.
\end{equation}
On the other hand, the above jump conditions, equilibrium, and
equality along the contact discontinuity lead to
\begin{equation}
f  = \frac{(\gamma_2-1)(4\gamma_2+3)}{(\gamma_{34}-1)(4\gamma_{34}+3)}\simeq \frac{4\gamma_2^2}{(\gamma_{34}-1)(4\gamma_{34}+3)}.
\end{equation}
For a relativistic reverse shock (RRS), i.e., $\gamma_{34} \gg 1$ or
$f \ll \eta^2$, we have $\gamma_{34}\simeq \eta/2\gamma_2 =
\eta^{1/2}f^{-1/4}/\sqrt{2}$,
$\gamma_2=\gamma_3\simeq\eta^{1/2}f^{1/4}/\sqrt{2}$. For a
non-relativistic (Newtonian) reverse shock (NRS), i.e. $\gamma_{34}
\simeq 1$ or $\eta^{2} \ll f$, we have $\gamma_{34}-1 \simeq
4\eta^2/7f$, $\gamma_2=\gamma_3\simeq\eta$.

The distance $dR$ over which the reverse shock front travels and the
length $dx$ of propagation of the reverse shock in the unshocked
shell, satisfy the following equation (see also Sari \& Piran 1995):
\begin{equation}
dR=\frac{dx}{\beta_4-\beta_3}\left(1-\frac{\gamma_4 n'_4}{\gamma_3
n'_3}\right),
\end{equation}
where the second term on the right hand of the above equation
reflects the shock compression of the fluid contained in the $dx$.
In terms of $f$, we get (Kobayashi 2000; Wu et al 2003)
\begin{equation}
dR = \alpha \eta \sqrt f dx,
\end{equation}
in which the coefficient
\begin{equation}
\alpha=\frac{1+2\gamma_3/\eta}{\sqrt{4(\gamma_3/\eta)^2+6\gamma_3/\eta+4}},
\end{equation}
where $\alpha\simeq1/2$ for RRS ($\gamma_3\ll\eta$) and
$\alpha\simeq3/\sqrt{14}$ for NRS ($\gamma_3\simeq\eta$), as given
in Sari \& Piran (1995). The increase of the electron number in the
shocked shell (region 3) corresponds to the decrease of the electron
number in the unshocked fireball shell (region 4), which reads
\begin{equation}
dN_3 = -dN_4 = 4\pi R^2 \eta n'_4 dx = 4\pi \alpha^{-1}R^2 f^{1/2}n_1 dR.
\end{equation}
The total number of electrons in the initial shell is $N_0 = E/\eta
{m_p}{c^2}$. So the reverse shock crossing radius $R_{\Delta}$ is
determined by
\begin{equation}
N_0 = \int_{0}^{R_{\Delta}} 4\pi \alpha^{-1}R^2 f^{1/2}n_1 dR.
\end{equation}
In the observer's frame we have $dR = 2{\Gamma ^2}cdT/(1+z)$, where
$T$ is the observer time, $\Gamma$ is the Lorentz factor of the
shock front. For an ultra-relativistic shock, the bulk Lorentz
factor of the fluid just behind the shock front is
$\gamma=\Gamma/\sqrt{2}$ (Blandford \& McKee 1976). In this paper,
we adopt the homogeneous-thin-shell approximation and assume that the
bulk Lorentz factor of the whole shell is $\gamma$. Here, we use $dR=
4{\gamma ^2}cdT/(1+z)$.

In general, we can work out the hydrodynamic evolution of the
reverse-forward shocks by the above equations and initial
conditions. Before we proceed to obtain analytical solutions for the
problem, we compare four characteristic radii, which have been
introduced to study this problem (Sari \& Piran 1995 for $k=0$; Wu
et al. 2003; Zou et al. 2005; Granot 2012 for $k=2$) as follows.

(1) The reverse shock crossing radius ${R_\Delta }$, which can be
approximated by
\begin{equation}
{R_\Delta }  \simeq \Delta \eta \sqrt f \simeq {\left( {\frac{{\Delta \;{l^{3 - k}}}}{{3 - k}}}
\right)^{\frac{1}{{4 - k}}}}.
\end{equation}

(2) The transition radius ${R_N}$, which is defined when the reverse
shock changes from Newtonion to relativistic ($f = {\eta ^2}$),
\begin{equation}
{R_N} \simeq {\left[ {\frac{{\;{l^{3 - k}}}}{{(3 - k)\,\Delta \,{\eta
^4}}}} \right]^{\frac{1}{{2 - k}}}}.
\end{equation}

(3) The spreading radius ${R_S}$, which is
\begin{equation}
{R_S} \simeq \Delta_0 \;{\eta ^2}.
\end{equation}
Taking into account the spreading effect, the width of the shell is
$\Delta\simeq\Delta_0+R/\eta^2$. For $R<R_S$,
$\Delta\simeq\Delta_0$; for $R>R_S$, $\Delta\simeq R/\eta^2$.

(4) The deceleration radius ${R_\eta }$, which is defined when the
mass of the swept-up ambient medium $M_{\rm sw}$ by the forward
shock equals $M_0/\eta$,
\begin{equation}
{R_\eta } = \frac{l}{{{\eta ^{\frac{2}{{3 - k}}}}}},
\end{equation}
where $M_0=E/\eta c^2$ is the initial mass of the fireball shell.

Therefore, we define
\begin{equation}
\xi  \equiv {\left( {\frac{l}{\Delta }}
\right)^{\frac{1}{2}}}\;{\eta ^{ - \frac{{4 - k}}{{3 - k}}}},
\end{equation}
so the four radii follow the relation
\begin{equation}
\frac{{{R_N}}}{{{\xi ^{\frac{2}{{2 - k}}}}}} \simeq {R_\eta } \simeq
{\xi ^{\frac{2}{{4 - k}}}}{R_\Delta } \simeq {\xi ^2}{R_S}.
\end{equation}

In the case of $\xi < 1$, the order of the four radii is ${R_N} <
{R_\eta } < {R_\Delta } < {R_S}$ ($0\leq k \leq 2$), or  ${R_\eta }
< {R_\Delta } \leq R_S \leq  {R_N}  $ ($2 < k < 3$). ${R_S} >
R_{\Delta}$ means that the radial spreading of the shell is
unimportant, and $\Delta = {\Delta _0}$. This is the so-called
``thick shell'' case, as the initial width of the shell is thick
enough so that the spreading can be neglected. In this case, $R_N <
R_{\Delta}$ means that the reverse shock is relativistic for $0 \leq k
\leq 2$. However, for $2 < k < 3$, $R_{\Delta} < R_N $ does not mean
that the reverse shock is Newtonian. For $2 < k < 3$, $f$ is
proportional to $R^{k-2}$, which is initially much smaller than
$\eta^2$. The evolution of an reverse shock for $2 < k < 3$ is thus
from initially relativistic to non-relativistic later. This is
because the ambient medium density drops steeply with radius. So
$R_{\Delta} < R_N $ in the case of $2 < k < 3$ indeed means that the
reverse shock is relativistic. In general, the reverse shock is
always relativistic for $\xi  < 1$.

In the case of $\xi > 1$, the order of the four radii is ${R_S} <
{R_\Delta } < {R_\eta } < {R_N}$ ($0\leq k < 2$) or  ${R_N} \leq R_S
\leq {R_\Delta } < {R_\eta }$ ($2 \leq k < 3$). ${R_S} < R_{\Delta}$
means that the radial spreading is important, and $\Delta \simeq
R/\eta^2$. This is the so-called ``thin shell'' case, because the initial
width of the shell is thin enough that the spreading is dominant.
In this case, we rewrite the expressions for the crossing radius and
transition radius, and obtain
\begin{equation}
{R_N} \simeq {R_\Delta} \simeq {R_\eta} \simeq l/\eta^{2/(3-k)}\simeq  {\xi^2}{R_S}, \;\; \xi >1.
\end{equation}
The above relation shows that the reverse shock becomes mildly
relativistic when it just crosses the shell. This can also be drawn
from $f \sim \eta^2$ at the crossing radius. Since $f\propto
R^{k-3}$ in this case, we can see that $f$ is a decreasing function
of $R$ for $k<3$, or $f$ is much larger than $\eta^2$ at a smaller
radius. In the following, in order to work out the analytical
solution, we treat the thin shell case by assuming that the reverse
shock is non-relativistic.


\subsection{The Thick Shell Case ($\xi < 1$)}
The reverse shock in the thick shell case can be assumed to be
relativistic. The density ratio in the thick shell case is
\begin{equation}
f = \frac{{{l^{3 - k}}}}{{(3 - k){R^{2 -
k}}\Delta_0 {\eta ^2}}}.
\end{equation}
The crossing radius is
\begin{equation}
R_{\Delta}=\left[\frac{(4-k)^2 l^{3-k}\Delta_0}{16(3-k)}\right]^{1/(4-k)},
\end{equation}
and the crossing time of the reverse shock in the observer's frame is
\begin{equation}
T_{\Delta}=(1+z)\int_{0}^{R_{\Delta}}\frac{dR}{4\gamma_3^2 c} = (1+z)\frac{\Delta_0}{4c}.
\end{equation}
At the crossing time, the bulk Lorentz factor of the shocked fluid
(both the shocked shell and shocked ambient medium) is
\begin{equation}
{\gamma_{3,\Delta}}= {\gamma_{2,\Delta}} \simeq \frac{1}{\sqrt{2}}
\eta^{1/2} f_{\Delta}^{1/4}= \left[2^k (3-k)
(4-k)^{2-k}\right]^{-1/2(4-k)}\left(\frac{l}{\Delta_0}\right)^{(3-k)/2(4-k)}.
\end{equation}

For the reverse shock, the relative Lorentz factor between the
shocked shell and un-shocked shell at $R_{\Delta}$ is
\begin{equation}
{\gamma_{34,\Delta}} \simeq  \frac{1}{2}
\frac{\eta}{\gamma_{3,\Delta}} = \left[2^{3k-8} (3-k)
(4-k)^{2-k}\right]^{1/2(4-k)} \eta
\left(\frac{l}{\Delta_0}\right)^{-(3-k)/2(4-k)}.
\end{equation}
The number density and pressure of the shocked shell at the crossing
time are
\begin{equation}
n'_{3,\Delta} \simeq \frac{8\gamma_{3,\Delta}^3 n_{1,\Delta}}{\eta}
=  \left[2^{24-k} (3-k)^{2k-3} (4-k)^{-(6+k)}\right]^{1/2(4-k)}
\frac{A}{\eta}
\left(l^{(3-2k)(3-k)}\Delta_0^{k-9}\right)^{1/2(4-k)},
\end{equation}
and $e'_{3,\Delta}\simeq \gamma_{34,\Delta}n'_{3,\Delta}m_p c^2$,
respectively. The total number of electrons at $R_{\Delta}$ is
$N_{3,\Delta}=N_0$.

For the forward shock, the number density and pressure of the
shocked surrounding medium at $R_{\Delta}$ are
\begin{equation}
n'_{2,\Delta} \simeq 4\gamma_{2,\Delta} n_{1,\Delta}  =
\left[2^{16+3k} (3-k)^{2k-1} (4-k)^{-(2+3k)}\right]^{1/2(4-k)} A
\left(l^{(3-k)(1-2k)}\Delta_0^{-(3+k)}\right)^{1/2(4-k)},
\end{equation}
and $e'_{2,\Delta}\simeq \gamma_{2,\Delta}n'_{2,\Delta}m_p c^2$,
respectively. We assume a pressure balance across the contact
discontinuity, $e'_2\simeq e'_3$, so we have $e'_{2,\Delta}\simeq
e'_{3,\Delta}$. The total number of electrons in the shocked ambient
medium at $R_{\Delta}$ is
\begin{equation}
N_{2,\Delta}=\frac{4\pi}{3-k}n_{1,\Delta}R_{\Delta}^3 =
\frac{4\pi}{3-k} A \left[\frac{(4-k)^2
l^{3-k}\Delta_0}{16(3-k)}\right]^{(3-k)/(4-k)}.
\end{equation}

\subsubsection{The Shocked Shell}

Before the reverse shock crosses the shell, the hydrodynamic
evolution of the reverse shock can be characterized by ($T \leq T_{\Delta}$)
\begin{equation}
{\gamma _3} = \gamma_{3,\Delta}\left(\frac{T}{T_{\Delta}}\right)^{-(2 - k)/2(4 - k)},
\;\; R=R_{\Delta}\left(\frac{T}{T_{\Delta}}\right)^{2/(4 - k)}, \;\; N_3 = N_{3,\Delta}\frac{T}{T_{\Delta}},
\end{equation}
and
\begin{equation}
{n'_3} = n_{3,\Delta}\left(\frac{T}{T_{\Delta}}\right)^{-(6 + k)/2(4 - k)},
\;\; p'_3 = p'_{3,\Delta}\left(\frac{T}{T_{\Delta}}\right)^{-(2 + k)/(4 - k)}.
\end{equation}

After the reverse shock crosses the shell, the shocked shell
temperature and pressure are very high. The hydrodynamics of the
shocked shell will be dominated by its adiabatic expansion. On the
other hand, since the shocked shell is located not too far from the
forward shock, it can be roughly regarded as the tail of the forward
shock and so it follows the Blandford-McKee solution (Kobayashi \& Sari
2000). Therefore, we assume ${\gamma _3} \propto {R^{\frac{{2k -
7}}{2}}}, {e'_3} \propto {R^{\frac{{4k - 26}}{3}}},{n'_3} \propto
{R^{\frac{{2k - 13}}{2}}}$, and $T \propto R/{\gamma _3^2c}$. So the
hydrodynamic evolution of the reverse shock after crossing the shell
is characterized by ($T > T_{\Delta}$)
\begin{equation}
{\gamma _3} = \gamma_{3,\Delta}\left(\frac{T}{T_{\Delta}}\right)^{(2k - 7)/4(4 - k)},
\;\; R=R_{\Delta}\left(\frac{T}{T_{\Delta}}\right)^{1/2(4 - k)}, \;\; N_3 = N_{3,\Delta},
\end{equation}
and
\begin{equation}
{n'_3} = n_{3,\Delta}\left(\frac{T}{T_{\Delta}}\right)^{(2k - 13)/4(4 - k)},
\;\; e'_3 = e'_{3,\Delta}\left(\frac{T}{T_{\Delta}}\right)^{(2k - 13)/3(4 - k)}.
\end{equation}

\subsubsection{The Shocked Surrounding Medium}

Before the reverse shock crosses the shell, the hydrodynamic
evolution of the forward shock can be characterized by ($T \leq T_{\Delta}$)
\begin{equation}
{\gamma _2} =
\gamma_{2,\Delta}\left(\frac{T}{T_{\Delta}}\right)^{-(2 - k)/2(4 -
k)}, \;\; R=R_{\Delta}\left(\frac{T}{T_{\Delta}}\right)^{2/(4 - k)},
\;\; N_2 = N_{2,\Delta}\left(\frac{T}{T_{\Delta}}\right)^{2(3 -
k)/(4 - k)},
\end{equation}
and
\begin{equation}
{n'_2} = n_{2,\Delta}\left(\frac{T}{T_{\Delta}}\right)^{-(2 + 3k)/2(4 - k)},
\;\; e'_2 = e'_{2,\Delta}\left(\frac{T}{T_{\Delta}}\right)^{-(2 + k)/(4 - k)}.
\end{equation}

After the reverse shock crosses the shell, the hydrodynamics of the
forward shock follows the Blandford-McKee self-similar solution.
Because most of the energy and mass are contained within $\sim
R/\gamma_2^2$, hereafter we adopt the uniform thin shell
approximation. The hydrodynamics of the forward shock for $T >
T_{\Delta}$ is thus characterized by ($T > T_{\Delta}$)
\begin{equation}
{\gamma _2} =
\gamma_{2,\Delta}\left(\frac{T}{T_{\Delta}}\right)^{-(3 - k)/2(4 -
k)}, \;\; R=R_{\Delta}\left(\frac{T}{T_{\Delta}}\right)^{1/(4 - k)},
\;\; N_2 =  N_{2,\Delta}\left(\frac{T}{T_{\Delta}}\right)^{(3 -
k)/(4 - k)},
\end{equation}
and
\begin{equation}
{n'_2} = n_{2,\Delta}\left(\frac{T}{T_{\Delta}}\right)^{-(3 + k)/2(4 - k)},
\;\; e'_2 = e'_{2,\Delta}\left(\frac{T}{T_{\Delta}}\right)^{-3/(4 - k)}.
\end{equation}

\subsection{The Thin Shell Case ($\xi>1$)}

The reverse shock in the thin shell case can be assumed to be
non-relativistic. Because the spreading effect is important in this
case, the width of the shell is $\Delta\simeq R/\eta^2$. The density
ratio is thus
\begin{equation}
f = \frac{1}{3-k}\left(\frac{l}{R}\right)^{3-k}.
\end{equation}
The crossing radius is
\begin{equation}
R_{\Delta}=\left[\frac{9(3-k)}{56}\right]^{1/(3-k)}\frac{l}{\eta^{2/(3-k)}},
\end{equation}
and the crossing time of the reverse shock in the observer's frame
is
\begin{equation}
T_{\Delta}=\left[\frac{9(3-k)}{14 \times 4^{4-k}}\right]^{1/(3-k)}
\frac{(1+z)l}{\eta^{2(4-k)/(3-k)}c},
\end{equation}
assuming $\gamma_2=\gamma_3\simeq\eta$ throughout the entire
duration of the reverse shock crossing the shell.

For the reverse shock, the relative Lorentz factor between the
shocked shell and un-shocked shell at $R_{\Delta}$ is
\begin{equation}
{\gamma_{34,\Delta}} \simeq  1+\frac{4\eta^2}{7f_{\Delta}} = 1 + \frac{9(3-k)^2}{98}.
\end{equation}
The number density and pressure of the shocked shell at the crossing
time are
\begin{equation}
n'_{3,\Delta} \simeq 7 f_{\Delta} n_{1,\Delta}  =   \left[\frac{2^9
7^{6-k}}{3^6 (3-k)^{6-k}}\right]^{1/(3-k)} A l^{-k} \eta^{6/(3-k)},
\end{equation}
and $e'_{3,\Delta}\simeq (\gamma_{34,\Delta}-1)n'_{3,\Delta}m_p
c^2$, respectively. The total number of electrons at $R_{\Delta}$ is
$N_{3,\Delta}=N_0$.

For the forward shock, the number density and pressure of the
shocked surrounding medium at $R_{\Delta}$ are
\begin{equation}
n'_{2,\Delta} \simeq 4\gamma_{2,\Delta} n_{1,\Delta}  =
\left[\frac{2^{6+k} 7^{k}}{9^{k} (3-k)^{k}}\right]^{1/(3-k)} A
l^{-k} \eta^{(3+k)/(3-k)},
\end{equation}
and $e'_{2,\Delta}\simeq \gamma_{2,\Delta}n'_{2,\Delta}m_p c^2$,
respectively. We assume a pressure balance across the contact
discontinuity, $e'_2\simeq e'_3$, so we have $e'_{2,\Delta}\simeq
e'_{3,\Delta}$. The total number of electrons in the shocked ambient
medium at $R_{\Delta}$ is
\begin{equation}
N_{2,\Delta}=\frac{4\pi}{3-k}n_{1,\Delta}R_{\Delta}^3 =
\frac{9\pi}{14} \frac{A l^{3-k}}{ \eta^{2}}.
\end{equation}

\subsubsection{The Shocked Shell}

Before the reverse shock crosses the shell, the hydrodynamic
evolution of the reverse shock can be characterized by ($T \leq
T_{\Delta}$)
\begin{equation}
{\gamma _3} \simeq \eta, \;\; R=R_{\Delta}\frac{T}{T_{\Delta}},
\;\; N_3 = N_{3,\Delta}\left(\frac{T}{T_{\Delta}}\right)^{(3-k)/2},
\end{equation}
and
\begin{equation}
{n'_3} = n_{3,\Delta}\left(\frac{T}{T_{\Delta}}\right)^{-3},  \;\;
e'_3 = e'_{3,\Delta}\left(\frac{T}{T_{\Delta}}\right)^{-k}.
\end{equation}

After the reverse shock crosses the shell, similar to the thick
shell case, the hydrodynamic evolution of the reverse shock is
characterized by ($T > T_{\Delta}$)
\begin{equation}
{\gamma _3} = \eta \left(\frac{T}{T_{\Delta}}\right)^{(2k - 7)/4(4 -
k)},  \;\; R=R_{\Delta}\left(\frac{T}{T_{\Delta}}\right)^{1/2(4 -
k)}, \;\; N_3 = N_{3,\Delta},
\end{equation}
and
\begin{equation}
{n'_3} = n_{3,\Delta}\left(\frac{T}{T_{\Delta}}\right)^{(2k -
13)/4(4 - k)},  \;\; e'_3 =
e'_{3,\Delta}\left(\frac{T}{T_{\Delta}}\right)^{(2k - 13)/3(4 - k)}.
\end{equation}

\subsubsection{The Shocked Surrounding Medium}

Before the reverse shock crosses the shell, the hydrodynamic
evolution of the forward shock can be characterized by ($T \leq
T_{\Delta}$)
\begin{equation}
{\gamma _2} \simeq \eta, \;\;  R=R_{\Delta}\frac{T}{T_{\Delta}}, \;\; N_2 =
N_{2,\Delta}\left(\frac{T}{T_{\Delta}}\right)^{3 - k},
\end{equation}
and
\begin{equation}
{n'_2} = n_{2,\Delta}\left(\frac{T}{T_{\Delta}}\right)^{-k},  \;\;
e'_2 = e'_{2,\Delta}\left(\frac{T}{T_{\Delta}}\right)^{-k}.
\end{equation}

After the reverse shock crosses the shell, the hydrodynamics of the
forward shock is similar to the case of the thick shell, which
follows the Blandford-McKee solution and can be described as ($T >
T_{\Delta}$)
\begin{equation}
{\gamma _2} = \eta \left(\frac{T}{T_{\Delta}}\right)^{-(3 - k)/2(4 -
k)},  \;\; R=R_{\Delta}\left(\frac{T}{T_{\Delta}}\right)^{1/(4 -
k)}, \;\; N_2 = N_{2,\Delta}\left(\frac{T}{T_{\Delta}}\right)^{(3 -
k)/(4 - k)},
\end{equation}
and
\begin{equation}
{n'_2} = n_{2,\Delta}\left(\frac{T}{T_{\Delta}}\right)^{-(3 + k)/2(4
- k)},  \;\; e'_2 =
e'_{2,\Delta}\left(\frac{T}{T_{\Delta}}\right)^{-3/(4 - k)}.
\end{equation}

\section{Emission from the Reverse-Forward Shocks}
We assume that the afterglow of a GRB is due to the synchrotron
radiation of relativistic electrons with a power-law energy
distribution, $N(\gamma'_e)d\gamma'_e = N_{\gamma'_e}
\gamma_e^{'-p}d\gamma'_e$ $(\gamma'_e>\gamma'_m)$, where $\gamma'_m$
is the minimum Lorentz factor of the shock-accelerated electrons,
and $p$ is the power-law index of the energy distribution. Assuming
the two fractions $\epsilon_e$ and $\epsilon_B$, then the energy
densities contained in the electrons and magnetic field are $U'_e =
\epsilon_e e'$ and $U'_B = B'^2/8\pi = \epsilon_B e'$, respectively.
The minimum Lorentz factor and cooling Lorentz factor of electrons
evolve as ${\gamma'_m} \propto \epsilon_e{p'}/{n'}$ and ${\gamma'_c}
\propto B^{'-2}\gamma^{-1}T^{-1}(1+Y)^{-1}$. The Compton parameter
$Y$ is the ratio of the radiation energy density to the magnetic
energy density. Cooling of electrons will change the energy
distribution of the electrons (Sari et al. 1998). If the cooling Lorentz
factor is smaller than the minimum Lorentz factor, then the energy
distribution is altered to $N(\gamma'_e)\propto \gamma_e^{'-2}$ for
$\gamma'_c\leq\gamma'_e\leq\gamma'_m$, and $N(\gamma'_e)\propto
\gamma_e^{'-(p+1)}$ for $\gamma'_m<\gamma'_e$. Otherwise, the energy
distribution is $N(\gamma'_e)\propto \gamma_e^{'-p}$ for
$\gamma'_m\leq\gamma'_e\leq\gamma'_c$, and $N(\gamma'_e)\propto
\gamma_e^{'-(p+1)}$ for $\gamma'_c<\gamma'_e$. The characteristic
frequency radiated by an electron with $\gamma'_e$ in the observer's
frame is $\nu \simeq 0.5 \gamma\gamma^{'2}_{e}\nu_L/(1+z)$, where
$\nu_L= q_e B'/2\pi m_e c$ is the Lamour frequency, where $q_e$ and $m_e$
are the charge and rest mass of an electron. Thus, the scaling laws
for the typical frequency, cooling frequency and peak flux density
of synchrotron radiation are $\nu_m \propto B'\;\gamma
\;{\gamma_m^{'2}} \propto {e^{'\frac{5}{2}}}\;\gamma \;{n^{'-2}}$,
${\nu_c} \propto 1/{B^{'3}}\gamma {T^2}(1+Y)^2 \propto
{e^{'-\frac{3}{2}}}\;{\gamma ^{ - 1}}\;{T^{ - 2}}(1+Y)^{-2}$, and
${F_{\nu ,\max }} =(1+z)\frac{s-1}{2}\frac{{N_e}P_{\nu,\rm max}}
{4\pi D_L^2} \propto {N_e}\;\gamma \;{e^{'\frac{1}{2}}}$,
respectively. ${N_e}$ is the total number of electrons responsible
for synchrotron radiation, where $D_L$ is the luminosity distance of
the source. The peak spectral power is $P_{\nu,\rm max}\simeq m_e
c^2 \sigma_T \gamma B'/1.5q_e$. We note that only a fraction
$(s-1)/2$ of the total electrons contribute to the radiation at
the peak frequency of $F_\nu$,  where $s=2$ for the fast cooling
case ($\nu'_c< \nu'_m$) and $s=p$ for the slow cooling case
($\nu'_m< \nu'_c$).

\subsection{Reverse Shock Emission}
We now consider the synchrotron emission from the shocked shell.
When the reverse shock crosses the shell, it heats the shell and
accelerates electrons to form a relativistic non-thermal
distribution in the shocked region. Although we investigate the
reverse shock emission in this paper, we will not pay too much
attention to this emission component. This is because the reverse
shock emission is rarely identified in GRBs - only a
very small fraction of GRBs have shown the reverse shock emission
component in their early light curves. However, tens of GRBs have been
identified with an afterglow onset feature at early times, which is
attributed to the forward shock emission. The evolution of the
typical frequency, cooling frequency, and peak flux density of the
reverse shock follows the dynamics and the properties of the downstream
medium, i.e., ${\nu _m} \propto e_3^{\prime \frac{5}{2}}\;{\gamma
_3}\;n_3^{\prime - 2}$, ${\nu _c} \propto e_3^{\prime -
\frac{3}{2}}\;\gamma _3^{ - 1}\;{T^{ - 2}}$, and $F_{\nu ,\max
}^{RS} \propto {N_{3,e}}\;{\gamma _3}\;e_3^{\prime \frac{1}{2}}$.

\subsubsection{The Thick Shell Case}
The reverse shock is relativistic in the thick shell case. The time
of the reverse shock crossing the thick shell is comparable to the
duration of GRB prompt emission, i.e., $T_{\Delta} \sim T_{90}$. The
typical frequency, cooling frequency and peak flux density of the
reverse shock at the reverse shock crossing time $T_\Delta$ are
(e.g., Sari \& Piran 1999; Waxman \& Draine 2000)
\begin{equation}
{\nu^{r}_{m,\Delta}} = q_e \eta^2 \epsilon_e^{r2}
\left(\frac{p-2}{p-1}\right)^2
\frac{m_p^3}{m_e^3}(1+z)^{-1}\left(\frac{\epsilon_B^{r} A}{8\pi
m_p}\right)^{1/2}\left[\frac{(4-k)^2 E T_{\Delta}}{16\pi A m_p c
(1+z)}\right]^{-k/2(4-k)},
\end{equation}
\begin{equation}
{\nu^{r}_{c,\Delta}} = \frac{9(4-k)^2\pi m_e
q_e}{2(1+Y^r)^2\sigma_T^2(1+z)(8\pi \epsilon_{B}^{r}A m_p)^{3/2}}
\left[\frac{(4-k)^2 E T_{\Delta}}{16\pi A m_p c
(1+z)}\right]^{(3k-4)/2(4-k)},
\end{equation}
and
\begin{equation}
F_{\nu ,\max, \Delta }^{RS} = \frac{(s-1)(1+z)^2 m_e \sigma_T (8\pi
\epsilon_{B}^{r}A m_p)^{1/2} E}{12(4-k) \pi m_p q_e \eta T_{\Delta}
D_L^2} \left[\frac{(4-k)^2 E T_{\Delta}}{16\pi A m_p c
(1+z)}\right]^{(2-k)/2(4-k)}.
\end{equation}

The scaling laws before and after the reverse shock crossing
$T_\Delta$ are
\begin{equation}
T < T_{\Delta}:\,\,{\nu^{r}_m} \propto {T^{ - \frac{k}{{4 - k}}}},{\nu^{r}_c}
\propto {T^{\frac{{3k - 4}}{{4 - k}}}},F_{_{\nu ,\max }}^{RS}
\propto {T^{\frac{{2 - k}}{{4 - k}}}},
\end{equation}
and
\begin{equation}
T > T_{\Delta}:\,\,{\nu^{r}_m} \propto {T^{\frac{{14k - 73}}{{12(4 -
k)}}}},{\nu^{r}_c} \propto {T^{\frac{{14k - 73}}{{12(4 -
k)}}}},F_{_{\nu ,\max }}^{RS} \propto {T^{\frac{{10k - 47}}{{12(4 -
k)}}}}.
\end{equation}
Due to the adiabatic cooling, the evolution ${\nu _c}$ ($\gamma_c$)
is assumed to be the same as ${\nu _m}$ ($\gamma_m$) after the
reverse shock crosses the shell (Kobayashi 2000).

\subsubsection{The Thin Shell Case}
In the thin shell case, the reverse shock is non relativistic, so it
is too weak to decelerate the shell effectively. The spreading of
the shell is significant in this case, so the time of the reverse
shock crossing the shell is much longer than the duration of GRB
prompt emission, i.e., $T_{\Delta} \gg T_{90}$.  The typical
frequency, cooling frequency and peak flux density of the reverse
shock at the reverse shock crossing time $T_\Delta$ are
\begin{equation}
{\nu^{r}_{m,\Delta}} =
\left[\frac{9^{12-5k}(3-k)^{24-10k}}{2^{9-6k}7^{24-9k}}\right]^{1/2(3-k)}
\epsilon_e^{r2}
\left(\frac{p-2}{p-1}\right)^2\frac{m_p^3}{m_e^3}\frac{q_e
\eta^{(6-k)/(3-k)}}{1+z}\left(\frac{\epsilon_B^{r} A}{\pi
m_p}\right)^{1/2}\left[\frac{E}{4\pi A m_p c^2}\right]^{-k/2(3-k)},
\end{equation}
\begin{equation}
{\nu^{r}_{c,\Delta}} = \left[\frac{9^{2+k} 7^{4-3k}}{2^{9+2k}
(3-k)^{8-6k} }\right]^{1/2(3-k)} \frac{\pi m_e}{\sigma_T^2(\pi
\epsilon_B^{r}m_p A)^{3/2}}\frac{q_e
\eta^{(4-3k)/(3-k)}}{(1+z)(1+Y^{r})^2}\left[\frac{E}{4\pi A m_p
c^2}\right]^{(3k-4)/2(3-k)},
\end{equation}
and
\begin{equation}
F_{\nu ,\max,\Delta }^{RS} = \left[\frac{2^{15-2k}
7^{k}}{3^{6}(3-k)^{2k}}\right]^{1/2(3-k)}
(s-1)(1+z)\eta^{3/(3-k)}A^{3/2}\left(\pi \epsilon_B^r
m_p\right)^{1/2}\frac{m_e c^3 \sigma_T}{q_e D_L^2}
\left[\frac{E}{4\pi A m_p c^2}\right]^{3(2-k)/2(3-k)}.
\end{equation}

The scaling laws before and after the reverse shock crossing time
are
\begin{equation}
T < {T_\Delta }:\,\,{\nu^{r}_m} \propto {T^{\frac{{12 -
5k}}{2}}},{\nu^{r}_c} \propto {T^{\frac{{3k - 4}}{2}}},F_{\nu ,\max
}^{RS} \propto {T^{\frac{{3 - 2k}}{2}}},
\end{equation}
and
\begin{equation}
T > {T_\Delta }:\,\,{\nu^{r}_m} \propto {T^{\frac{{14k - 73}}{{12(4 -
k)}}}},{\nu^{r}_c} \propto {T^{\frac{{14k - 73}}{{12(4 -
k)}}}},F_{_{\nu ,\max }}^{RS} \propto {T^{\frac{{10k - 47}}{{12(4 -
k)}}}}.
\end{equation}
We assume that the equation of state of the shocked shell is mildly
relativistic so it can be regarded as the tail of the forward shock,
satisfying the Blandford-McKee self-similar solution (see Kobayashi
2000 for an alternative treatment). However, since the reverse shock
emission is usually not observed (suppressed by the forward shock)
in the thin shell case, this assumption is unimportant.

\subsection{Forward Shock Emission}

Most of GRBs have been detected with the forward shock emission at
early times. A significant fraction of GRBs have shown the afterglow
onset feature (Liang et al. 2010). In this paper, we focus on the
forward shock emission, and investigate the effect of environments.
The evolution of the typical frequency, cooling frequency and peak
flux density of the forward shock follows the dynamics and the
properties of the downstream medium, i.e., ${\nu _m} \propto e_2^{\prime
\frac{5}{2}}\;{\gamma _2}\;n_2^{\prime - 2}$, ${\nu _c} \propto
e_2^{\prime - \frac{3}{2}}\;\gamma _2^{ - 1}\;{T^{ - 2}}$, and
$F_{_{\nu ,\max }}^{FS} \propto {N_{2,e}}\;{\gamma _2}\;e_2^{\prime
\frac{1}{2}}$.

\subsubsection{The Thick Shell Case}
The two characteristic frequencies and peak flux density at the
reverse shock crossing time $T_{\Delta}$ in the thick shell case are
\begin{equation}
{\nu^{f}_{m,\Delta}} = \frac{1}{2^{7/2}(4-k)\pi}(\epsilon_e^{f})^{2}
(\epsilon_B^{f})^{1/2}  \left(\frac{p-2}{p-1}\right)^{2}
\left(\frac{m_p}{m_e}\right)^2 \frac{q_e}{m_e c}{\left[
\frac{(1+z)E}{(cT_\Delta)^3} \right]^{\frac{1}{2}}},
\end{equation}
\begin{equation}
{\nu^{f}_{c,\Delta}} = \frac{9\pi (4-k)^2}{4^{k/(4-k)}}\frac{q_e
m_e}{(1+z) \sigma_T^2 (1+Y^f)^2 (8\pi \epsilon_B^f A m_p)^{3/2}}
\left[\frac{(4-k)^2 E T_\Delta}{4\pi A m_p c
(1+z)}\right]^{(3k-4)/2(4-k)},
\end{equation}
and
\begin{equation}
F_{\nu,\max,\Delta }^{FS} =  \frac{(4-k)}{3\pi(3-k)
2^{(16-5k)/(4-k)}} \frac{(1+z)(s-1)m_e \sigma_T E}{q_e m_p D_L^2}
(8\pi \epsilon_B^f A m_p c^2)^{1/2} \left[\frac{(4-k)^2 E
T_\Delta}{4\pi A m_p c (1+z)}\right]^{-k/2(4-k)}.
\end{equation}

The scaling law before and after the reverse shock crossing time are
\begin{equation}
T < T_{\Delta}:\,\,{\nu^{f}_m} \propto {T^{ - 1}},{\nu^{f}_c} \propto {T^{\frac{{3k
- 4}}{{4 - k}}}},F_{\nu ,\max }^{FS} \propto {T^{\frac{{4 - 2k}}{{4
- k}}}},
\end{equation}
and
\begin{equation}
T > T_{\Delta}:\,\,{\nu^{f}_m} \propto {T^{ - \frac{3}{2}}},{\nu^{f}_c} \propto
{T^{\frac{{3k - 4}}{{2(4 - k)}}}},F_{\nu ,\max }^{FS} \propto
{T^{\frac{{ - k}}{{2(4 - k)}}}}.
\end{equation}

As we can see, for the forward shock emission in the thick shell
case, the evolution of ${\nu_m}$ is independent of $k$, and hence
does not depend on the distribution of the ambient medium.

\subsubsection{The Thin Shell Case}

The two characteristic frequencies and peak flux density at the
reverse shock crossing time $T_{\Delta}$ in the thin shell case are
\begin{equation}
{\nu^{f}_{m,\Delta}} = \left[\frac{2^{3+2k} 7^{k}}{9^{k}
(3-k)^{2k}}\right]^{1/2(3-k)}  (\epsilon_e^{f})^2
\left(\frac{p-2}{p-1}\right)^2\frac{m_p^3}{m_e^3}\frac{q_e
\eta^{3(4-k)/(3-k)}}{1+z}\left(\frac{\epsilon_B^{f} A}{\pi
m_p}\right)^{1/2}\left[\frac{E}{4\pi A m_p c^2}\right]^{-k/2(3-k)},
\end{equation}
\begin{equation}
{\nu^{f}_{c,\Delta}} = \left[\frac{9^{2+k} 7^{4-3k}}{2^{9+2k}
(3-k)^{8-6k} } \right]^{1/2(3-k)} \frac{\pi m_e}{\sigma_T^2(\pi
\epsilon_B^{f}m_p A)^{3/2}}\frac{q_e
\eta^{(4-3k)/(3-k)}}{(1+z)(1+Y^{f})^2}\left[\frac{E}{4\pi A m_p
c^2}\right]^{(3k-4)/2(3-k)},
\end{equation}
and
\begin{equation}
F_{\nu ,\max,\Delta }^{FS} = \left[\frac{9^{3-2k}
(3-k)^{6-4k}}{2^{3-4k} 7^{6-3k}}
\right]^{1/2(3-k)}(s-1)(1+z)\eta^{k/(3-k)}A^{3/2}(\pi \epsilon_B^f
m_p)^{1/2} \frac{m_e c^3 \sigma_T}{q_e D_L^2}{\left[
{\frac{{E}}{{4\pi A{m_p}{c^2}}}} \right]^{\frac{{6 - 3k}}{{2(3 -
k)}}}}.
\end{equation}

The scalings law before and after the reverse shock crossing time are
\begin{equation}
T < {T_\Delta }:\,\,{\nu^{f}_m} \propto {T^{ - \frac{k}{2}}},{\nu^{f}_c}
\propto {T^{\frac{{3k - 4}}{2}}},F_{\nu ,\max }^{FS} \propto
{T^{\frac{{6 - 3k}}{2}}},
\end{equation}
and
\begin{equation}
T > {T_\Delta }:\,\,{\nu^{f}_m} \propto {T^{ - \frac{3}{2}}},{\nu^{f}_c}
\propto {T^{\frac{{3k - 4}}{{2(4 - k)}}}},F_{\nu ,\max }^{FS}
\propto {T^{\frac{{ - k}}{{2(4 - k)}}}}.
\end{equation}

For $T>T_\Delta$, the forward shock enters the Blandford-McKee phase
either in the thin shell case or in the thick shell case. So the
hydrodynamics and temporal evolution of the characteristic
frequencies and peak flux density are the same in both cases after
the reverse shock crosses the shell. The theoretical
flux density before the reverse shock crosses the shell is
\begin{equation}
  \label{eq:A_k_nu}
  F_\nu ^{FS} (T<T_{\Delta})\propto \left\{
  \begin{array}{ll}
   {T^{ \frac{{8 - 2k - kp}}{4}}}{\nu ^{ - \frac{p}{2}}},\quad \nu
        > \max\left\{ {\nu _c^f,\nu _m^f} \right\} \\
   {T^{\frac{{12 - 5k - kp}}{4}}}{\nu ^{ - \frac{{p - 1}}{2}}},\quad
        \nu _m^f < \nu  < \nu _c^f\\
   {T^{\frac{{8 - 3k}}{4}}}\;{\nu ^{ - \frac{1}{2}}}\;.\quad \quad
       \;\nu _c^f < \nu  < \nu _m^f
  \end{array}
  \right.
\end{equation}

The theoretical flux density after the reverse shock crosses the shell is
\begin{equation}
  F_\nu ^{FS}(T>T_{\Delta})\propto \left\{
  \begin{array}{ll}
   {T^{ - \frac{{3p - 2}}{4}}}{\nu ^{ - \frac{p}{2}}},\quad \nu  >
       \max\left\{{\nu _c^f,\nu _m^f} \right\}\\
   {T^{ - \frac{{12p - 3kp - 12 + 5k}}{{4(4 - k)}}}}{\nu ^{ - \frac{{p
      - 1}}{2}}},\quad \nu _m^f < \nu  < \nu _c^f\\
   {T^{ - \frac{1}{4}}}\;{\nu ^{ - \frac{1}{2}}}\;.\quad \quad \;\nu
      _c^f < \nu  < \nu _m^f
  \end{array}
  \right.
\end{equation}
Figure 1 and 2 present theoretical light curves of the forward shock
emission at early times in the thin shell case for $k=1$.

\section{Case Study}
In this paper, we investigate the temporal evolution of the dynamics and
emission of these two shocks in a stratified medium with a power-law
density distribution when both thick-shell and thin-shell cases are
considered. The crossing time
$T_\bigtriangleup\sim\bigtriangleup_0/c$ is comparable to the GRB
duration time for the thick shell, while for the thin shell, the
crossing time $T_\bigtriangleup$ is larger than the GRB duration.
The observed peak time of early afterglow onset is typically larger
than the GRB duration in a statistical sense (Liang et al. 2010; Li
et al. 2012). Therefore, these early onset peaks can be explained as
the forward shock emission in the thin-shell case, and the peak time
of the onset peak is interpreted as the reverse shock crossing time.
Recently, Liang et al. (2013) estimated the values of $k$ for a
sample of early optical afterglow onset peaks by assuming $\nu _m^f
< \nu < \nu _c^f $. They took $k$ to be free for the rising phase,
but assumed $k=0$ for the decay phase. They found that $k$ is
generally less than 2 and the typical value of $k$ is $\sim1$. In
this paper, we consider a general power-law distribution of the
ambient medium density during the whole afterglow phase and
calculate the hydrodynamic evolution of forward-reverse shocks in
both the thick-shell and thin-shell cases. This is different from
Liang et al. (2013), as mentioned above. We do not consider the
case of $\nu < \left\{ {\nu _c^f,\nu _m^f} \right\}$, because it is
unlikely in optical and X-ray spectra. Synchrotron self-absorption
can be neglected in the X-ray emission, and may also be unimportant
most of the time for optical afterglows. So for simplicity, we do no
consider this effect in this paper.  We select 19 GRBs as a
sample to determine their $k$ values. Most of our sample are the
same as Liang et al. (2013). In the following, we take three
well-observed optical afterglows as example cases to test the
forward shock model discussed in this paper (see Fig. 3). The results of the
remaining GRBs are shown in Fig. 4 and Table 1.

\subsection{GRB 060605}

GRB 060605 is a relatively faint gamma-ray burst, that was detected
by {\it Swift}/BAT, with a redshift of $z = 3.773$ (Ferrero et al 2009).
From Sato et al. (2006), the burst in the 15-350 keV band had a
duration of $T_{90} = 15 \pm 2$ s. According to the traditional
$T_{90}$ classification method, GRB 060605 belongs to a long
duration burst. Because long bursts are widely believed to originate
from the collapse of massive stars, the circumburst medium of GRB
060605 might have been a stellar wind environment. The peak time
of this optical onset is $t_p=399.1\pm13.0$ s, while the rising and
decaying indices are $\alpha_1=0.90\pm0.09$ and
$\alpha_2=1.17\pm0.05$, respectively (Rykoff et al. 2009). Ferrero et al. (2009),
studied the broad-band spectrum of
the afterglow of GRB 060605 at $t=0.07$ days and obtained a spectral
index $\beta_o=1.04\pm0.05$. The correction for Galactic extinction
at the $R_c$-band was considered. We consider both $\nu  >
\max\left\{ {\nu _c^f,\;\nu_m^f} \right\}$ and $\nu _c^f > \nu > \nu
_m^f $ to interpret the spectral index $\beta_o=1.04\pm0.05$. Thus
we have two possible values for the power-law index of energy
distribution.

(1) $\nu  > \max\left\{ {\nu _c^f,\;\nu_m^f} \right\}$. In this
case, $p=2\beta_0=2.08\pm0.10$. The value of $p$ can also be derived
from the decay index, i.e., $p=(4\alpha_2+2)/3=2.23\pm0.17$, which
is consistent with that derived from the optical spectrum. The
theoretical rising index is $\alpha_1=(8-2k-kp)/4$ (see Eq. 67), so
$k=(8-4\alpha_1)/(2+p)=1.08\pm0.11$. The values $k$ and $p$ are both
reasonable. We thus apply the $\nu  > \max\left\{ {\nu
_c^f,\;\nu_m^f} \right\}$ case of the forward shock model to fit
this optical peak, adopting $k=1.08\pm0.11$ and $p=2.08\pm0.10$.

(2) $\nu _m^f < \nu < \nu _c^f $. In this case,
$p=2\beta_0+1=3.08\pm0.10$. The theoretical decaying index
$\alpha_2=(12p-3kp+5k-12)/(16-4k)$ (see Eq. 68). Therefore, we
obtain $k=(12p-16\alpha_2-12)/(3p-4\alpha_2-5)\sim-14.2$ with the
observed decaying index $\alpha_2=1.17\pm0.05$. This value of $k$ is
not reasonable. Thus the model with $\nu _m^f < \nu < \nu _c^f $
cannot explain the optical onset peak of GRB 060605.

From the two cases discussed above, we find that only the $\nu  >
\max\left\{ {\nu _c^f,\;\nu_m^f} \right\}$ case could be applied to
explain the optical afterglow onset of GRB 060605. We derive
$k=1.08\pm0.11$ and $p=2.08\pm0.10$. Fig.3 shows our model fitting
to the observed afterglow of GRB 060605. We can see that the medium
density profile $n\propto R^{-1.1}$ is required to fit the data of
GRB 060605. This implies that the circumburst medium of GRB 060605
is neither a homogenous interstellar medium with $k=0$ nor a typical
stellar wind environment with $k=2$, as previously assumed.

There is a total of seven physical parameters in our model, i.e. $E$,
$A (R_0)$, $\eta$, $k$, $p$, $\varepsilon_B$, and $\varepsilon_e$.
However, there are not enough observational conditions in GRB 060605
to derive the exact values of these parameters. We can only
constrain the range of these parameters with available conditions.
The values of $k$ and $p$ are estimated above for GRB 060605, so we
can fix $k=1.1$ and $p=2.1$ in the following calculation. Then we
obtain
\begin{equation}
{\nu^{f}_{m,\Delta}} = 1.41\times 10^{15} \varepsilon _{e,-1}^{2}\,\varepsilon _{B,-1}^{1/2}E_{53}^{-0.29}\eta_2^{4.58} R_{0,17}^{0.87}\,{\rm Hz},
\end{equation}
\begin{equation}
{\nu^{f}_{c,\Delta}} = 1.69\times 10^{14} \varepsilon _{B,-1}^{-3/2}(1+Y^f)^{-2}E_{53}^{-0.18}\eta_2^{0.37} R_{0,17}^{-1.45}\,{\rm Hz},
\end{equation}
\begin{equation}
F_{\nu ,\max, \Delta }^{FS} =  5.56\times 10^{-3} \varepsilon _{B,-1}^{1/2}E_{53}^{0.71}\eta_2^{0.58} R_{0,17}^{0.87}\,{\rm Jy}.
\end{equation}
Since the $\nu
> \max\left\{ {\nu _c^f,\;\nu_m^f} \right\}$ case is applied
to explain the optical onset peak of GRB 060605, we get two constraints, i.e., $\nu >
\nu_m^f$ and $\nu > \nu_c^f$, where $\nu=4.29\times10^{14}$ Hz is
the optical frequency. The constraints are shown as follows:

(1) The initial isotropically kinetic energy $E = (1 - {\eta _\gamma })/{\eta _\gamma
}\;{E_{\gamma ,iso}}$, where ${\eta _\gamma }$ is the radiation
efficiency of GRBs. The initial energy is $E\sim {\rm a \; few} \times
10^{53}$ erg as ${E_{\gamma ,iso,52}}=2.8\pm0.5$ (Li et
al. 2012; also see Ferrero et al. 2009) and ${\eta _\gamma }\sim0.2$.

(2) $\nu > \nu_m^f$. From Eq. (69), we
obtain ${R_{0,17}} < 0.26 E_{53}^{0.33}\varepsilon _{e,-1}^{-2.30}\,\varepsilon _{B,-1}^{-0.58}\,{\eta_2
^{-5.27}}$.

(3) $\nu > \nu_c^f$. From Eq. (70), we
obtain $  R_{0,17}\,> 0.53\varepsilon
_{B,-1}^{-1.04}\,\eta_2^{0.25}\,E_{53}^{-0.13}(1+Y^f)^{-1.38}$.

(4) The crossing time $T_\Delta=t_p$. From Eq. (34), we obtain
$R_{0,17}\, = 0.34 \,{E_{53}^{0.91}}{\eta_2 ^{-5.27}}$.

(5) The peak flux density of the optical onset $F_{\nu , p} \sim
3\times 10^{-3}$ Jy. From Eqs. (69) - (71), we obtain
\begin{equation}
F_{\nu, \Delta}^{FS} =  6.71\times 10^{-3} \varepsilon
_{e,-1}^{1.10}\varepsilon
_{B,-1}^{0.03}(1+Y^f)^{-1}E_{53}^{0.46}\eta_2^{3.28}
R_{0,17}^{0.62}\,{\rm Jy},
\end{equation}
for  $\nu > \max\left\{ {\nu _c^f,\;\nu_m^f} \right\}$. So we have
$R_{0,17}\, \sim 0.27 \,\varepsilon _{e,-1}^{-1.77}\varepsilon
_{B,-1}^{-0.04}(1+Y^f)^{1.61}{E_{53}^{-0.74}}{\eta_2 ^{-5.29}}$.

The allowed parameter values should satisfy the above constraints
(2) - (5). Combining the constraints (2) and (5), we have
$E_{53}>\varepsilon _{e,-1}^{0.5}\varepsilon
_{B,-1}^{0.5}(1+Y^f)^{1.5}$. Combining constraints (4) and (5),
we have $E_{53}=0.87\varepsilon _{e,-1}^{-1.1}\varepsilon
_{B,-1}^{-0.02}(1+Y^f)^{0.98}$. Combining constraints (2) and
(4), we have $E_{53} < 0.75 \varepsilon _{e,-1}^{-2.3}\varepsilon
_{B,-1}^{-0.58}$. Combining constraints (3) and (4), we have
$\eta_2 < 0.92 E_{53}^{0.19}\varepsilon
_{B,-1}^{0.19}(1+Y^f)^{0.25}$. From the above analysis, we can see
that the initial Lorentz factor $\eta\sim 100$, which is insensitive
to other parameters. In Fig.3, we fit the optical data of GRB 060605
by adopting the following parameter values, $k=1.08\pm0.11$,
$p=2.08\pm0.10$, $R_0=1\times10^{17}$ cm, $E=8\times10^{53}$ ergs,
$\eta=120$, $\varepsilon_B=0.2$ and $\varepsilon_e=0.02$.

\subsection{GRB 081203A}

GRB 081203A was detected and located by {\it Swift}/BAT with a
duration of $T_{90} = 294 \pm 71$\,s (Ukwatta et al. 2008). Optical
spectroscopic observation led to the measurement of the redshift
$z=2.05 \pm 0.01$ (Kuin et al. 2009). The peak time of this
afterglow onset is $T_{p} = 367.1 \pm 0.8$ s, the rise and
decay slopes are $\alpha_1=2.20\pm0.01$ and $\alpha_2=1.49\pm0.01$,
respectively (Kuin et al. 2009). The ultraviolet spectrum of this
GRB was observed with the index $\beta_o=0.90\pm0.01$ for the early
time. We also consider the two following scenarios $\nu  >
\max\left\{ {\nu _c^f,\;\nu_m^f} \right\}$ and $\nu _c^f > \nu > \nu
_m^f $ to determine the environment of this GRB.

(1) $\nu  > \max\left\{ {\nu _c^f,\;\nu_m^f} \right\}$. In this
case, we get $k=-0.24\pm0.01$ and $p=3.00\pm0.01$ for the
rising/decaying indices (see Eq.67, 68). This value of $k$ is not
reasonable, so this model is not suitable for GRB 081203A.

(2) $\nu _m^f < \nu < \nu _c^f $. In this case,
$p=2\beta_0+1=2.80\pm0.02$. We get $k=0.40\pm0.01$ and
$p=2.91\pm0.01$ with the rising-decaying indices. The values $k$ and
$p$ are both reasonable. In the following, we use this model to
constrain the other physical parameters for this GRB.

The two characteristic frequencies and peak flux density at the
reverse shock crossing time could be calculated with the above
derived $k=0.40$ and $p=2.91$,
\begin{equation}
{\nu^{f}_{m,\Delta}} = 3.56\times 10^{16} \varepsilon
_{e,-1}^{2}\,\varepsilon _{B,-1}^{1/2}E_{54}^{-0.077}\eta_2^{4.15}
R_{0,17}^{0.23}\,\,{\rm Hz},
\end{equation}
\begin{equation}
{\nu^{f}_{c,\Delta}} = 8.16\times 10^{13} \varepsilon
_{B,-1}^{-3/2}(1+Y^f)^{-2}E_{54}^{-0.54}\eta_2^{1.08}
R_{0,17}^{-0.38}\,\,{\rm Hz},
\end{equation}
\begin{equation}
F_{\nu ,\max, \Delta }^{FS} =  0.204 \varepsilon
_{B,-1}^{1/2}E_{54}^{0.92}\eta_2^{0.15} R_{0,17}^{0.23}\,\,{\rm Jy}.
\end{equation}
Since the $\nu _m^f < \nu < \nu
_c^f $ case is applied
to explain the optical onset peak of GRB 081203A, we get two constraints, i.e., $\nu >
\nu_m^f$ and $\nu_c^f > \nu$, where $\nu=4.29\times10^{14}$ Hz is
the optical frequency. The constraints are shown as follows:

(1) $E = (1 - {\eta _\gamma })/{\eta _\gamma
}\;{E_{\gamma ,iso}}$.The initial energy is $E\sim {\rm a \; few} \times
10^{54}$ erg as ${E_{\gamma ,iso,53}}=1.7\pm0.4$ erg (Li et al. 2012).

(2) $\nu > \nu_m^f$. From Eq. (69), we obtain ${R_{0,17}} <
4.53\times 10^{-9} E_{54}^{0.33}\varepsilon
_{e,-1}^{-8.69}\,\varepsilon _{B,-1}^{-2.17}\,{\eta_2 ^{-18.04}}$.

(3) $\nu < \nu_c^f$. From Eq. (70), we
obtain $  R_{0,17}\,< 0.013\varepsilon
_{B,-1}^{-3.95}\,\eta_2^{2.84}\,E_{54}^{-1.42}(1+Y^f)^{-5.26}$.

(4) The crossing time $T_\Delta=t_p$. From Eq. (34), we obtain
$R_{0,17}\, = 8.02 \,{E_{54}^{0.40}}{\eta_2 ^{-18}}$.

(5) The peak flux density of the optical onset $F_{\nu , p} \sim
2.6\times 10^{-2}$ Jy. From Eqs. (73) - (75), we get
\begin{equation}
R_{0,17}\, \sim 8.86\times 10^{-7} \,\varepsilon
_{e,-1}^{-4.22}\varepsilon _{B,-1}^{-2.17}{E_{54}^{-1.89}}{\eta_2
^{-9.09}}.
\end{equation}

The allowed parameter values should satisfy the above constraints
(2) - (5). Combining constraints (2) and (5), we have $\eta_2 <
0.556 E_{54}^{0.25}\varepsilon _{e,-1}^{-0.5}$. Combining
constraints (3) and (5), we have $\eta_2 > 0.45
E_{54}^{-0.04}\varepsilon _{e,-1}^{-0.35}\varepsilon
_{B,-1}^{0.15}(1+Y^f)^{0.44}$. Combining constraints (4) and
(5), we have $\eta_2 \sim 5.93 E_{54}^{0.25}\varepsilon
_{e,-1}^{0.47}\varepsilon _{B,-1}^{0.24}$. In Fig. 3, we fit the
optical data of GRB 081203A by adopting the following parameter
values, $k=0.40\pm0.01$, $p=2.91\pm0.01$, $R_0=1\times10^{17}$ cm,
$E=2\times10^{54}$ erg, $\eta=120$, $\varepsilon_B=0.01$ and
$\varepsilon_e=0.01$.

\subsection{XRF 071031}

The early light curve of the optical/near-infrared afterglow of the
X-Ray Flash (XRF) 071031 at $z=2.05$ with a duration of $T_{90} =
180 \pm 10$ s (Stamatikos et al. 2007, Kr{\"u}hler et al. 2009a)
shows a slow increase with flux $\propto T^{0.634\pm0.002}$ before
the peak time $T_{p} = 1018.6 \pm 1.6$ s. After the peak time, the
lightcurve decays with $T^{-0.845\pm0.001}$. The optical afterglow
spectral index is $\beta_o=0.9\pm0.1$.

(1) $\nu  > \max\left\{ {\nu _c^f,\;\nu_m^f} \right\}$. In this
case, $p=2\beta_0=1.8\pm0.2$. The value of $p$ can also be derived
from the decay index, i.e., $p=(4\alpha_2+2)/3=1.793\pm0.001$, which
is consistent with that derived from the optical spectrum. The
theoretical rising index is $\alpha_1=(8-2k-kp)/4$ (see Eq. 67), so
$k=(8-4\alpha_1)/(2+p)=1.440\pm0.001$. The values of $k$ and $p$ are
both reasonable. We thus apply the $\nu  > \max\left\{ {\nu
_c^f,\;\nu_m^f} \right\}$ case of the forward shock model to fit
this optical lightcurve, adopting $k=1.440\pm0.001$ and
$p=1.793\pm0.001$\footnote{For a flat energy distribution of
electrons with $p<2$, most of the energy of electrons is deposited in
electrons with minimal Lorentz factors, for details see Dai \& Cheng
(2001). Here we assume that the energy distribution of shock
injected electrons has a broken power law form, as introduced in Li
\& Chevalier (2001), thus the calculation of $\nu_m$ is the same as
in Sari et al. (1998).}.

(2) $\nu _m^f < \nu < \nu _c^f $. In this case,
$p=2\beta_0+1=2.8\pm0.2$. The theoretical decay index
$\alpha_2=(12p-3kp+5k-12)/(16-4k)$ (see Eq. 68). Therefore, we
obtain $k=(12p-16\alpha_2-12)/(3p-4\alpha_2-5)\sim404$ with the
observed decay index $\alpha_2=0.845\pm0.001$. This value of $k$ is
unreasonable. Thus the model with $\nu _m^f < \nu < \nu _c^f $
cannot explain the optical lightcurve of GRB 071031.

For $k=1.440$ and $p=1.793$, the two characteristic frequencies and
the peak flux density at the reverse shock crossing time are
\begin{equation}
{\nu^{f}_{m,\Delta}} = 2.2\times 10^{16} \varepsilon
_{e,-1}^{2}\,\varepsilon _{B,-1}^{1/2}E_{53}^{-0.46}\eta_2^{4.92}
R_{0,17}^{1.38}\,\,{\rm Hz},
\end{equation}
\begin{equation}
{\nu^{f}_{c,\Delta}} = 9.65\times 10^{13} \varepsilon
_{B,-1}^{-3/2}(1+Y^f)^{-2}E_{53}^{0.1}\eta_2^{-0.21}
R_{0,17}^{-2.31}\,\,{\rm Hz},
\end{equation}
\begin{equation}
F_{\nu ,\max, \Delta }^{FS} =  0.01 \varepsilon
_{B,-1}^{1/2}E_{53}^{0.54}\eta_2^{0.92} R_{0,17}^{1.38}\,\,{\rm Jy}.
\end{equation}
Since the $\nu  >
\max\left\{ {\nu _c^f,\;\nu_m^f} \right\}$ case is applied
to explain the optical lightcurve of GRB 071031, we get two constraints, i.e., $\nu >
\nu_m^f$ and $\nu > \nu_c^f$, where $\nu=4.29\times10^{14}$ Hz is
the optical frequency. The constraints are shown as follows:

(1) $E = (1 - {\eta _\gamma })/{\eta _\gamma
}\;{E_{\gamma ,iso}}$.The initial energy is $E\sim {\rm a \; few} \times
10^{53}$ erg as ${E_{\gamma ,iso,52}}=3.9\pm0.6$ erg (Li et al. 2012).

(2) $\nu > \nu_m^f$. From Eq. (69), we obtain ${R_{0,17}} < 0.058
E_{53}^{0.33}\varepsilon _{e,-1}^{-1.45}\,\varepsilon
_{B,-1}^{-0.36}\,{\eta_2 ^{-3.57}}$.

(3) $\nu > \nu_c^f$. From Eq. (70), we
obtain $  R_{0,17}\,> 0.52\varepsilon
_{B,-1}^{-0.65}\,\eta_2^{-0.09}\,E_{53}^{0.04}(1+Y^f)^{-0.87}$.

(4) The crossing time $T_\Delta=t_p$. From Eq. (34), we obtain
$R_{0,17}\, = 0.092 \,{E_{53}^{0.69}}{\eta_2 ^{-3.56}}$.

(5) The peak flux density of the optical onset $F_{\nu , p} \sim
1.66\times 10^{-4}$ Jy. From Eqs. (73) - (75), we get
\begin{equation}
R_{0,17}\, \sim 1.74\times 10^{-3} \,\varepsilon
_{e,-1}^{-1.03}\varepsilon _{B,-1}^{0.064}{E_{53}^{-0.53}}{\eta_2
^{-3.59}(1+Y^f)^{1.29}}.
\end{equation}

The allowed parameter values should satisfy the above constraints
(2) - (5). Combining the constraints (2) and (5), we have
$E_{53}>0.017\varepsilon _{e,-1}^{0.49}\varepsilon
_{B,-1}^{0.49}(1+Y^f)^{1.5}$. Combining constraints (4) and (5),
we have $E_{53}\sim 0.039\varepsilon _{e,-1}^{-0.84}\varepsilon
_{B,-1}^{0.05}(1+Y^f)^{1.06}$. Combining constraints (3) and
(4), we have $\eta_2 < 0.61 E_{53}^{0.19}\varepsilon
_{B,-1}^{0.19}(1+Y^f)^{0.25}$. In Fig. 3, we fit the optical data of
GRB 071031 by adopting the following parameter values,
$k=1.440\pm0.001$, $p=1.793\pm0.001$, $R_0=2\times10^{16}$ cm,
$E=5\times10^{53}$ erg, $\eta=90$, $\varepsilon_B=0.2$ and
$\varepsilon_e=0.02$.

\section{Discussion}

We have investigated the hydrodynamic evolution of a fireball
in both thick-shell and thin-shell cases, and considered
reverse-forward shocks in each case. According to the standard
fireball model, the reverse shock is initially non-relativistic for
the thin shell case, which is consistent with most of the onset
observations. If the GRB ejecta is highly magnetized ($\sigma\gg1$),
then the reverse shock will be significantly suppressed, and hence the
forward shock evolution will also be altered. Although observations
suggest that in some GRBs the ejecta is likely magnetized, the
degree of magnetization is usually $\sigma<$ a few at the radius
when the ejecta begins to decelerate. For simplicity we assume the
ejecta has no magnetization ($\sigma=0$) in this paper. For early
afterglows from GRB ejecta with non-negligible magnetization, please
see, e.g., Zhang, Kobayashi \& M{\'e}sz{\'a}ros (2003) and Zhang \&
Kobayashi (2005). Our paper aims to present analytical solutions
for the reverse-forward shock hydrodynamics and emission. In our
numerical fit to some GRB afterglow onset, we neglect the curvature
effect. The curvature effect, or more strictly speaking, the
equal-arrival-time-surface effect, has a minor effect on the
rise/decay slope of GRB afterglows.

A large number of multi-waveband afterglows have been detected since
the launch of {\em Swift}. The observations show that the optical and
X-ray afterglows of some bursts have different temporal properties.
A question thus arises: do afterglows at different wavebands have
the same origin? Here we analyze GRB 060605 as an example to discuss
this question. The smooth optical afterglow of this burst is assumed
to have been produced by the forward shock when the fireball was
decelerated by a circumburst medium in the $\nu  > \max\left\{ {\nu
_c^f,\;\nu_m^f} \right\}$ case. Figs. 3 shows the X-ray lightcurve
(diamonds) detected by {\em
Swift}\footnote{http://www.swift.ac.uk/xrt\_curves/}. The X-ray
lightcurve consists of three power-law segments with two break times
$t_{b1}=210\pm30$ s and $t_{b2}=7510\pm410$ s, which could be
described with a smoothly broken double power-law (Liang et al.
2008; Ferrero et al. 2009). The first segment decays quickly with
temporal index $\alpha_{\rm I}=2.19\pm0.42$, followed by a plateau
phase with $\alpha_{\rm II}=0.34\pm0.03$, then the third segment
starts with $\alpha_{\rm III}=1.89\pm0.07$ (Godet et al. 2006;
Ferrero et al. 2009). These properties have been summarized in a
canonical X-ray afterglow lightcurve scenario (Zhang et al. 2006;
Nousek et al. 2006): an initial steep decay followed by a shallow
decay phase, a normal decay, a post-jet break component and with
some erratic X-ray flares. The plateau phase of this burst is
currently understood as being due to ongoing energy injection. One
reasonable scenario is a fast rotating pulsar/magnetar as the the
central engine, which spins down through magnetic dipole radiation
(Dai \& Lu 1998b, c; Zhang \& M\'esz\'aros 2001; Dai 2004; Fan \& Xu
2006; Dai \& Liu 2012). There are also some flares after the prompt
GRB phase (Burrows et al. 2005; Falcone et al. 2006), which is
generally considered to be due to long-lasting central engine
activity (e.g., Fan \& Wei 2005; Dai et al. 2006; Ioka et al. 2005).
Thus, it is reasonable to assume that the X-ray afterglow of an
external forward shock is suppressed by the internal plateau
emission and X-ray flares for GRB 060605, as in Fig. 3. However, we
note that the X-ray afterglow at late times is likely dominated by
the forward shock emission. The other X-ray afterglows of our
sample also show these (or part) emission properties.

\section{Conclusions}
In this paper, we have investigated the evolution of the dynamics
and emission of the forward-reverse shocks in the circumburst
environment with general density distribution ${n_1} = A{R^{ - k}}$
by considering thick- and thin- shell cases. The optical afterglow
with one smooth onset peak at early times is usually attributed to
an external shock when the fireball is decelerated by a circumburst
medium. Long-duration GRBs may originate from the collapse of
massive stars and their ambient medium may be stellar winds. We
can infer the GRB circumburst medium from the rise and decay
features of the early onset peak (see Eqs. 67 and 68). We applied
our model to 19 GRBs, and found their $k$ values are in the range of
0.4-1.4, with a typical value of $k\sim1$ (see Fig. 5). This implies
that the circumburst medium of those GRBs is neither the ISM
($k=0$) nor a typical stellar wind ($k=2$). This could show a new
mass-loss evolution of the progenitor of this GRB, that is, the mass
loss rate $\dot{M}$ and/or the wind velocity $v_w$ are varied at
late times of the evolution of a massive star.

\section*{Acknowledgments}
We thank the anonymous referee for constructive suggestions.
We also thank En-Wei Liang, Xiang-Yu Wang, Yong-Feng Huang, Fa-Yin
Wang, Ruo-Yu Liu and Xuan Ding for useful comments and helps. This
work was supported by the National Basic Research Program of China
(grant No. 2014CB845800 and 2013CB834900) and the National Natural
Science Foundation of China (grant No. 11033002). XFW acknowledges
support by the One-Hundred-Talents Program and the Youth Innovation
Promotion Association of Chinese Academy of Sciences.

\clearpage
\begin{figure}
  \begin{center}
  \centerline{ \hbox{ \epsfig
  {figure=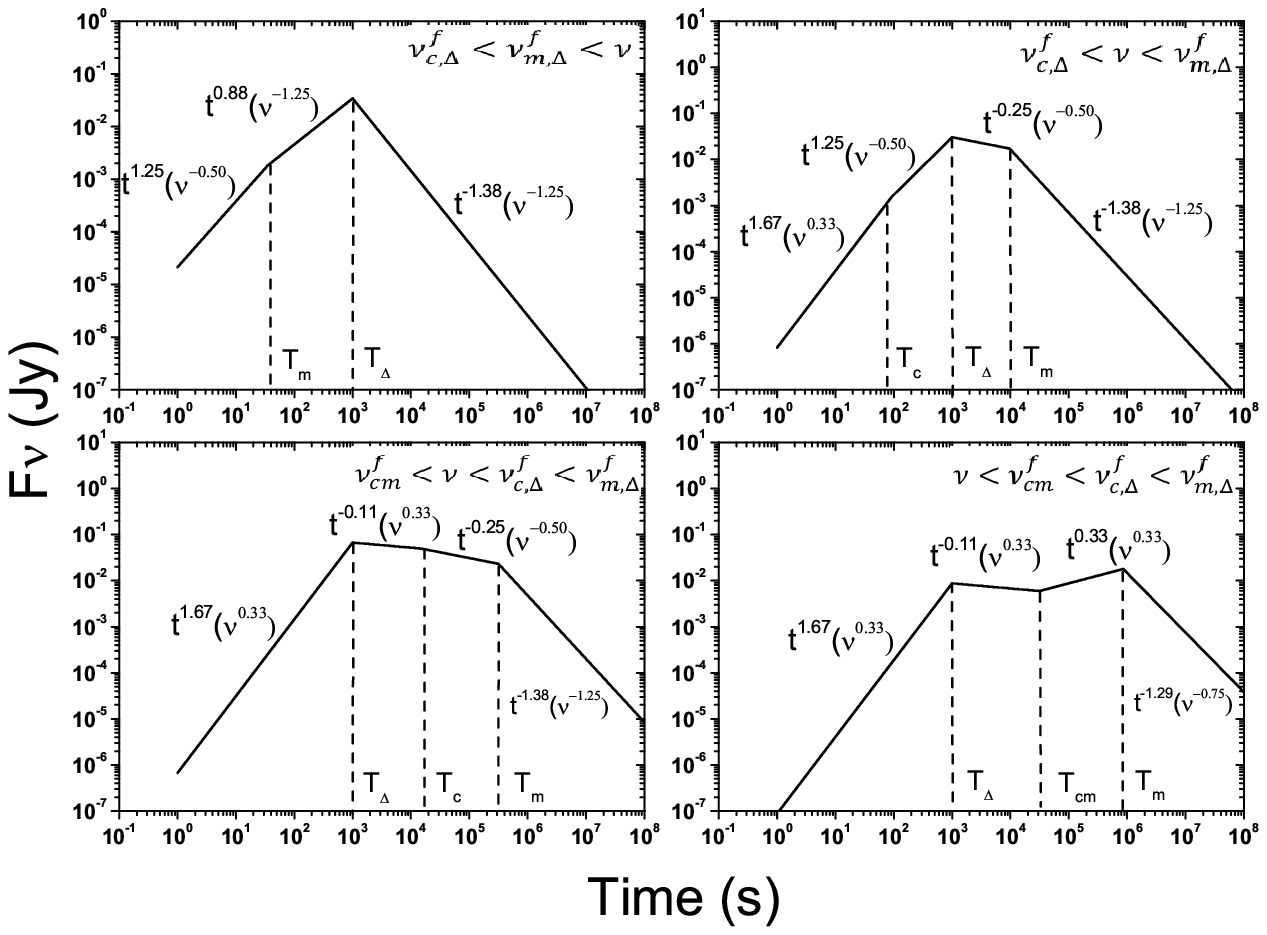,width=5.5in,height=3.8in,angle=0}}}
\caption{Characteristic light curves in the thin-shell fast-cooling
regime for $k=1$. $T_c$ and $T_m$ are the times when $\nu_c$ and
$\nu_m$ pass the observing frequency $\nu$, respectively. $T_{\rm
cm}$ is the time when $\nu_c=\nu_m$. The parameters of lightcurves
are $E=10^{52}$ erg, $\eta=100$, $z=1$, $R_0=10^{17}$ cm, $p=2.5$,
and $\epsilon _B^f=\epsilon _e^f=0.1.$}
  \end{center}
  \end{figure}

\clearpage
\begin{figure}
  \begin{center}
 \centerline{ \hbox{ \epsfig{figure=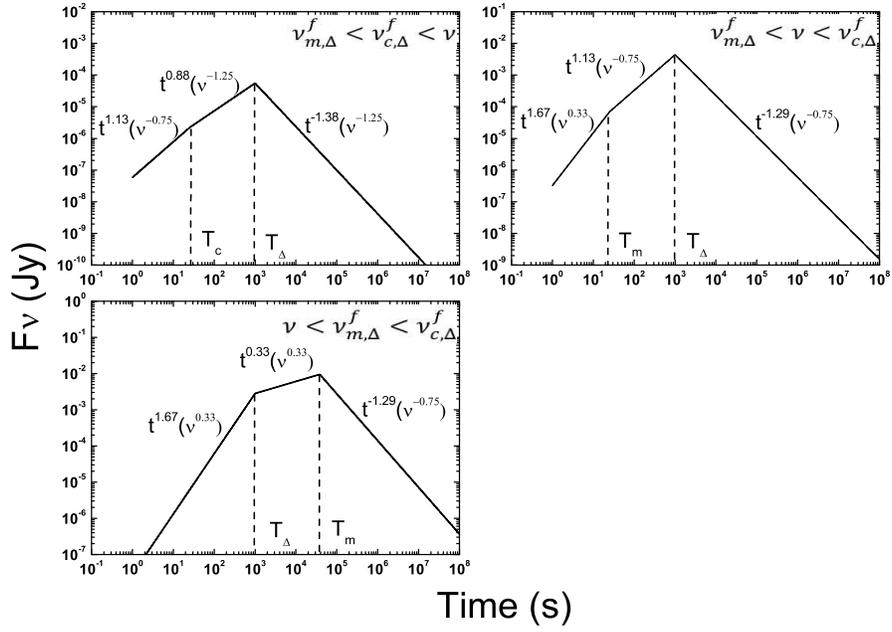,width=5.5in,height=3.8in,angle=0}}}
\caption{Characteristic light curves in the thin-shell slow-cooling
regime for $k=1$. The parameters of lightcurves are $E=10^{52}$ erg,
$\eta=100$, $z=1$, $R_0=10^{17}$ cm, $p=2.5$, and $\epsilon
_B^f=\epsilon _e^f=0.01.$}
  \end{center}
  \end{figure}

\newpage
\begin{figure*}
\includegraphics[angle=0,scale=0.5]{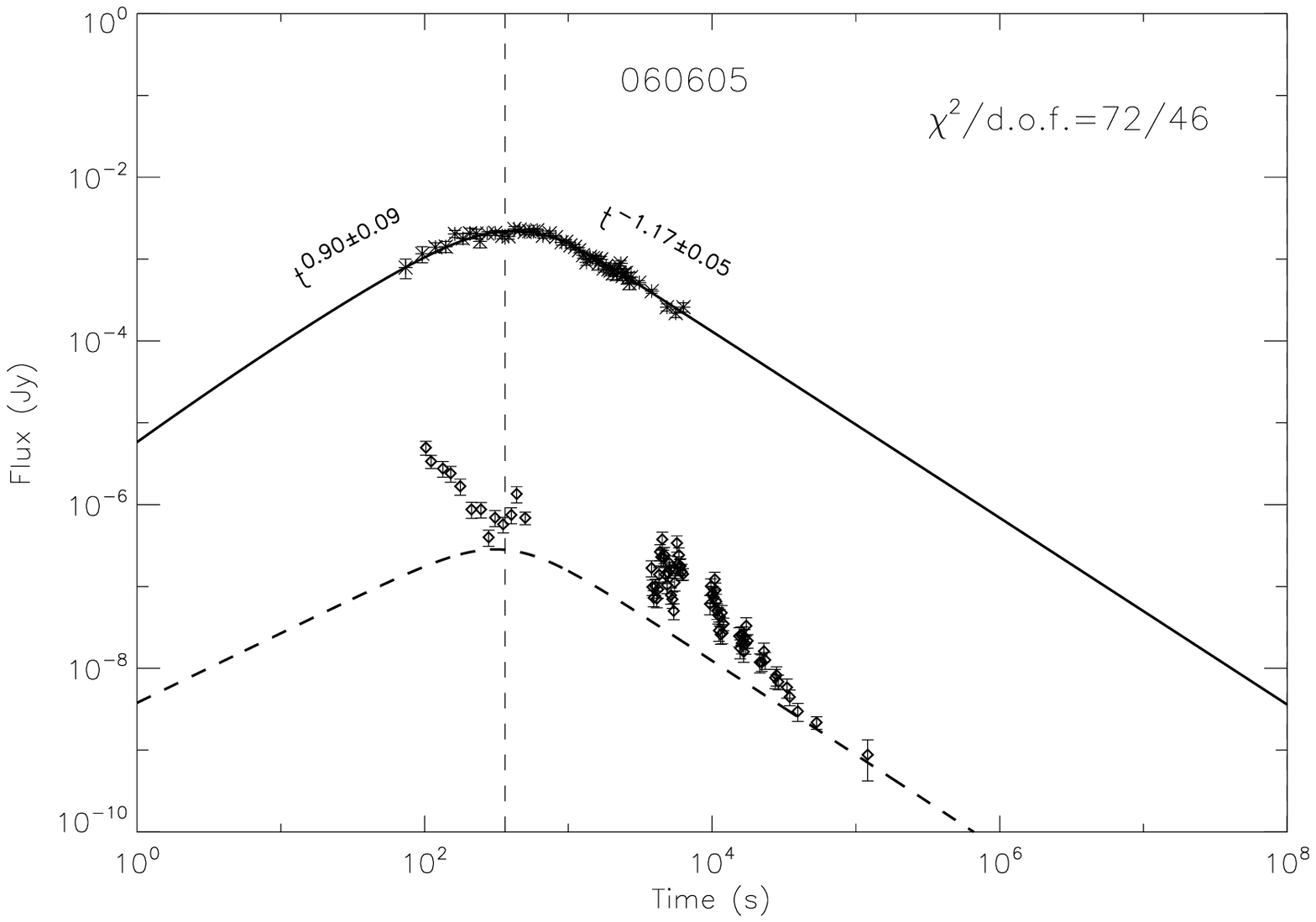}
\includegraphics[angle=0,scale=0.5]{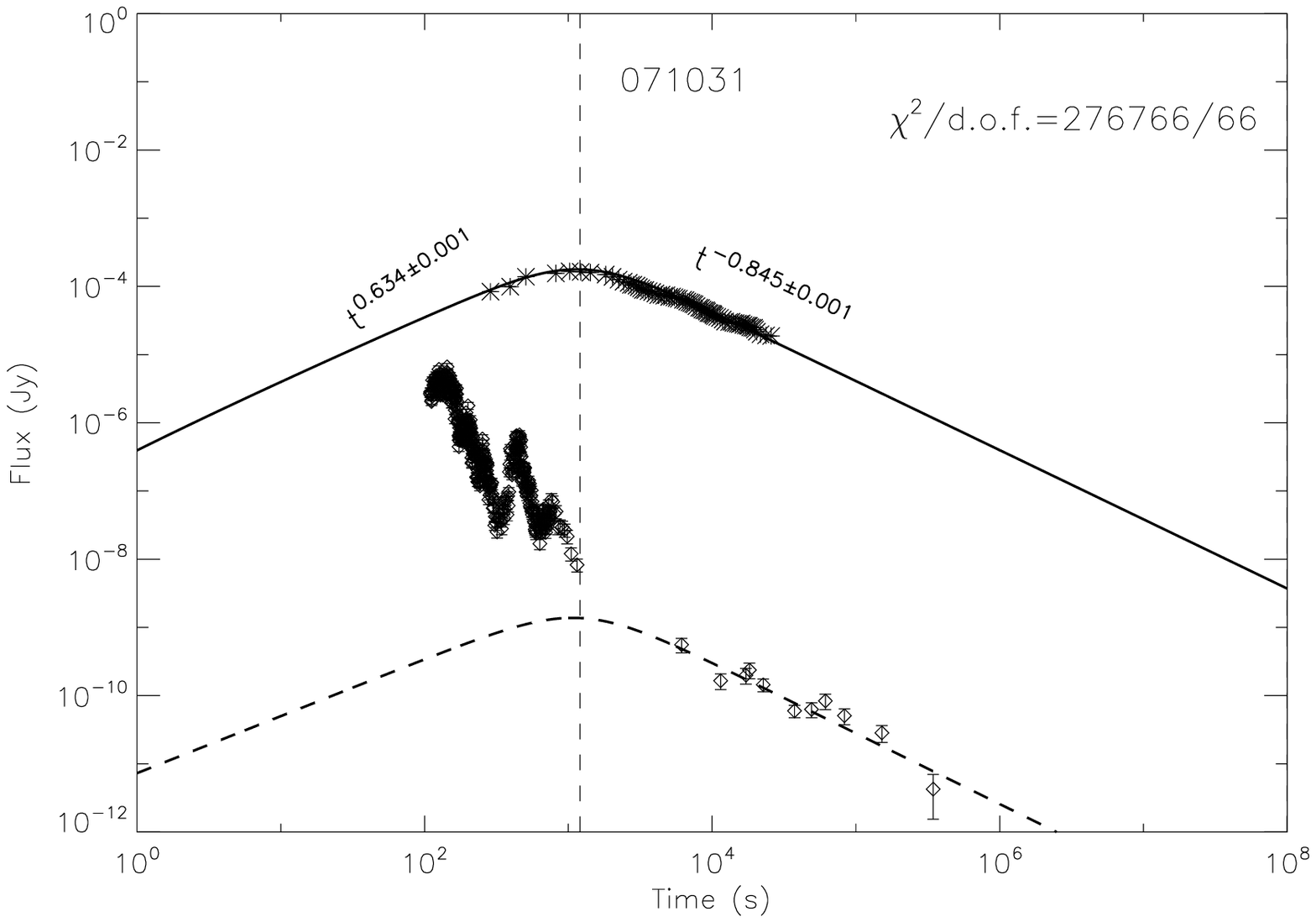}
\includegraphics[angle=0,scale=0.5]{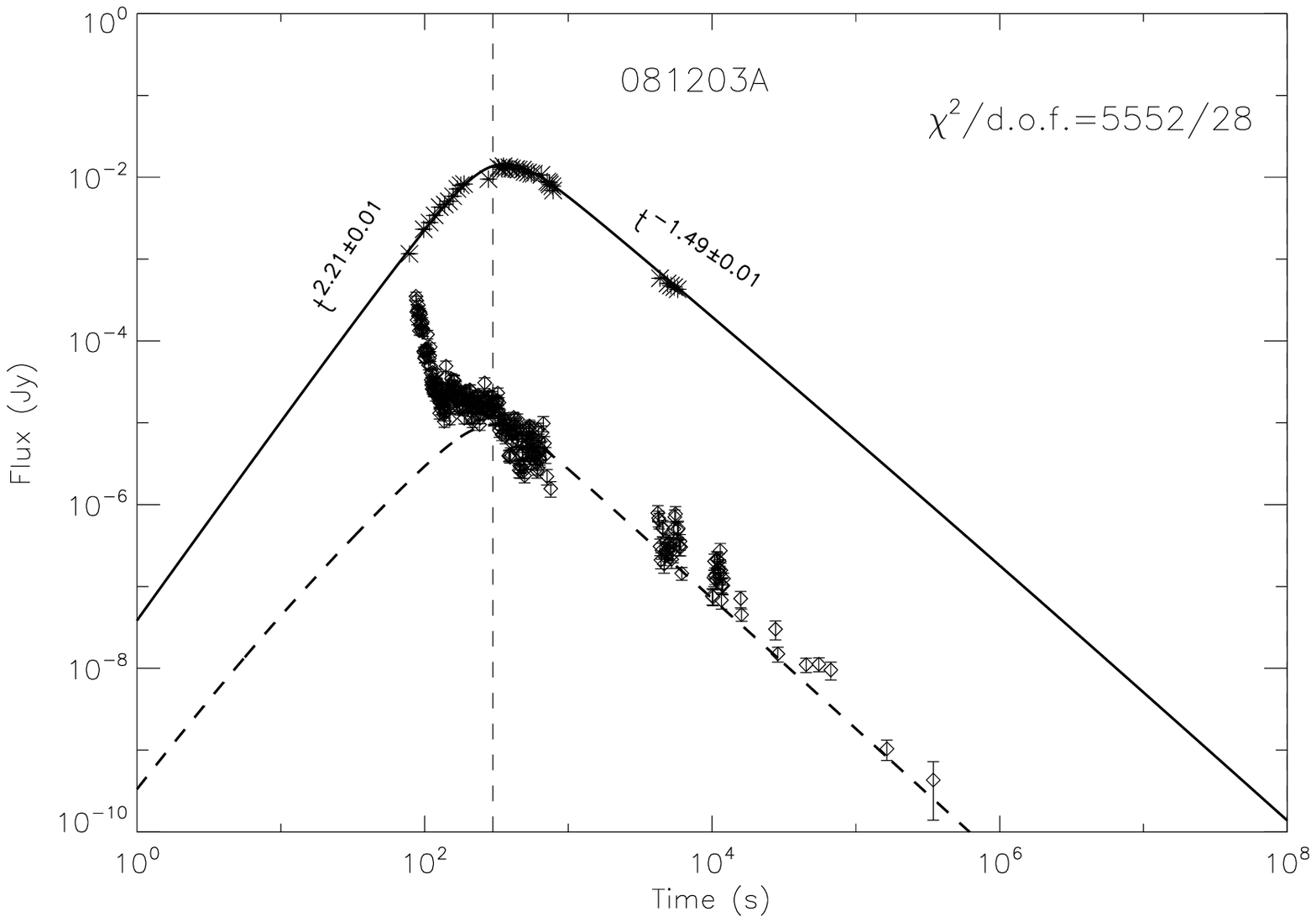}
\caption{The three typical GRBs of the sample(Optical light curve --
stars and X-ray light curve --diamonds). The $k$ values of GRBs
071031, 060605, 081203A are about 0.4, 1.0, 1.4, respectively. The
solid smooth line is the fit to the optical data, while the dashed
line corresponds to theoretical X-ray emission from the forward
shock. The vertical dashed line represents the peak time of the
lightcurve. The observed X-ray flux for GRB 071031, shown in this
figure, has been divided by 100.}
\end{figure*}

\clearpage
\begin{figure*}

\includegraphics[angle=0,scale=0.3]{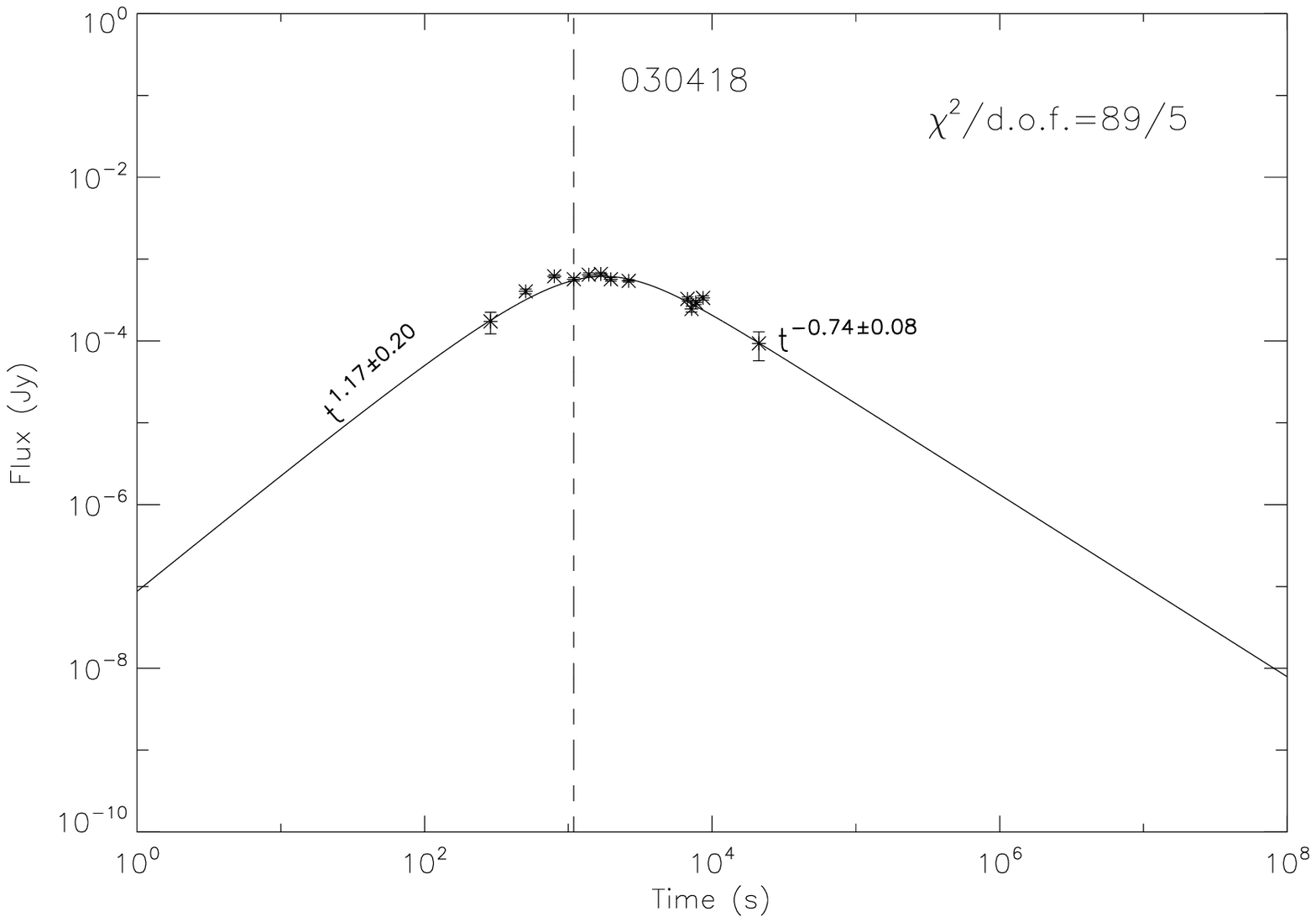}
\includegraphics[angle=0,scale=0.3]{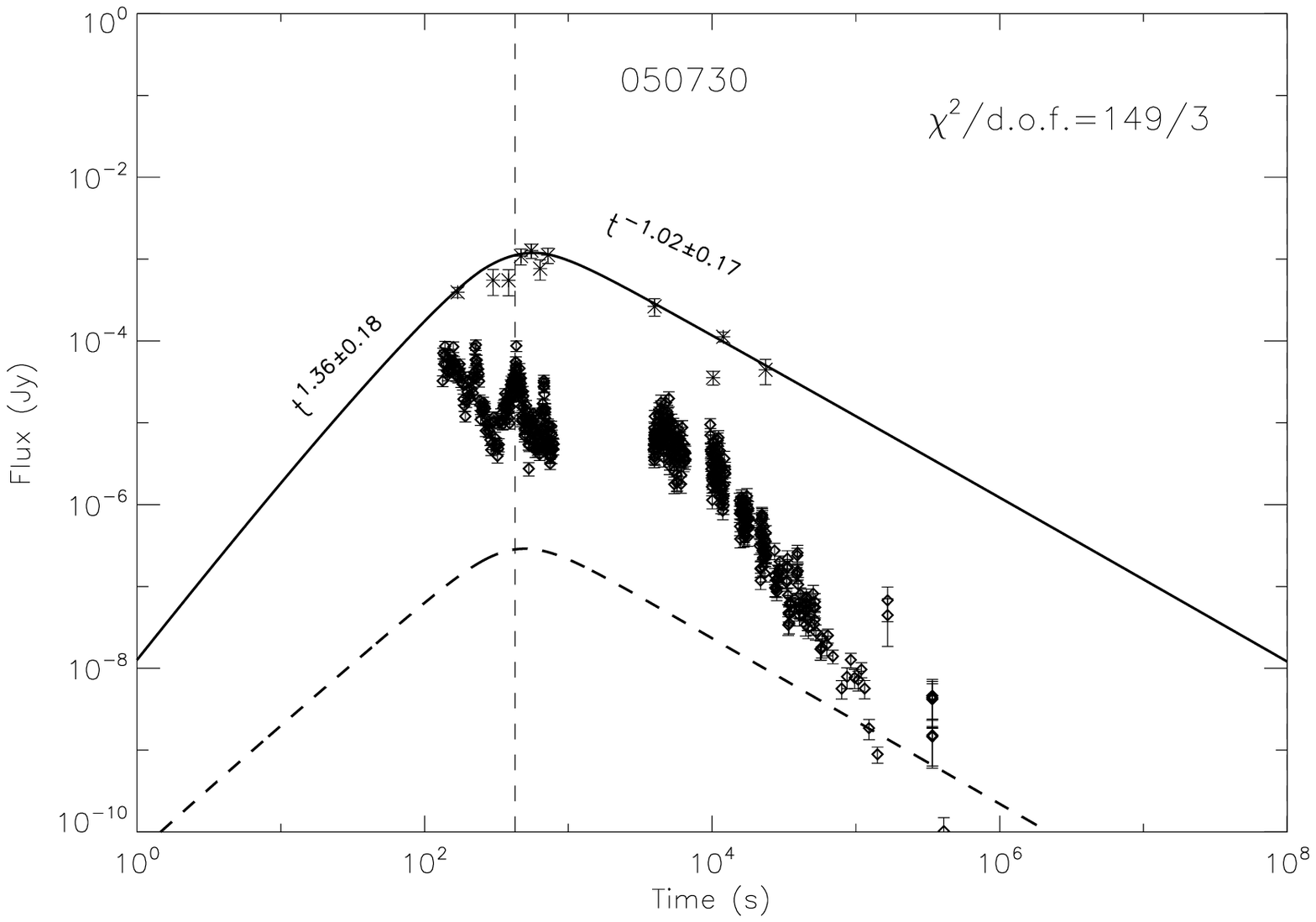}
\includegraphics[angle=0,scale=0.3]{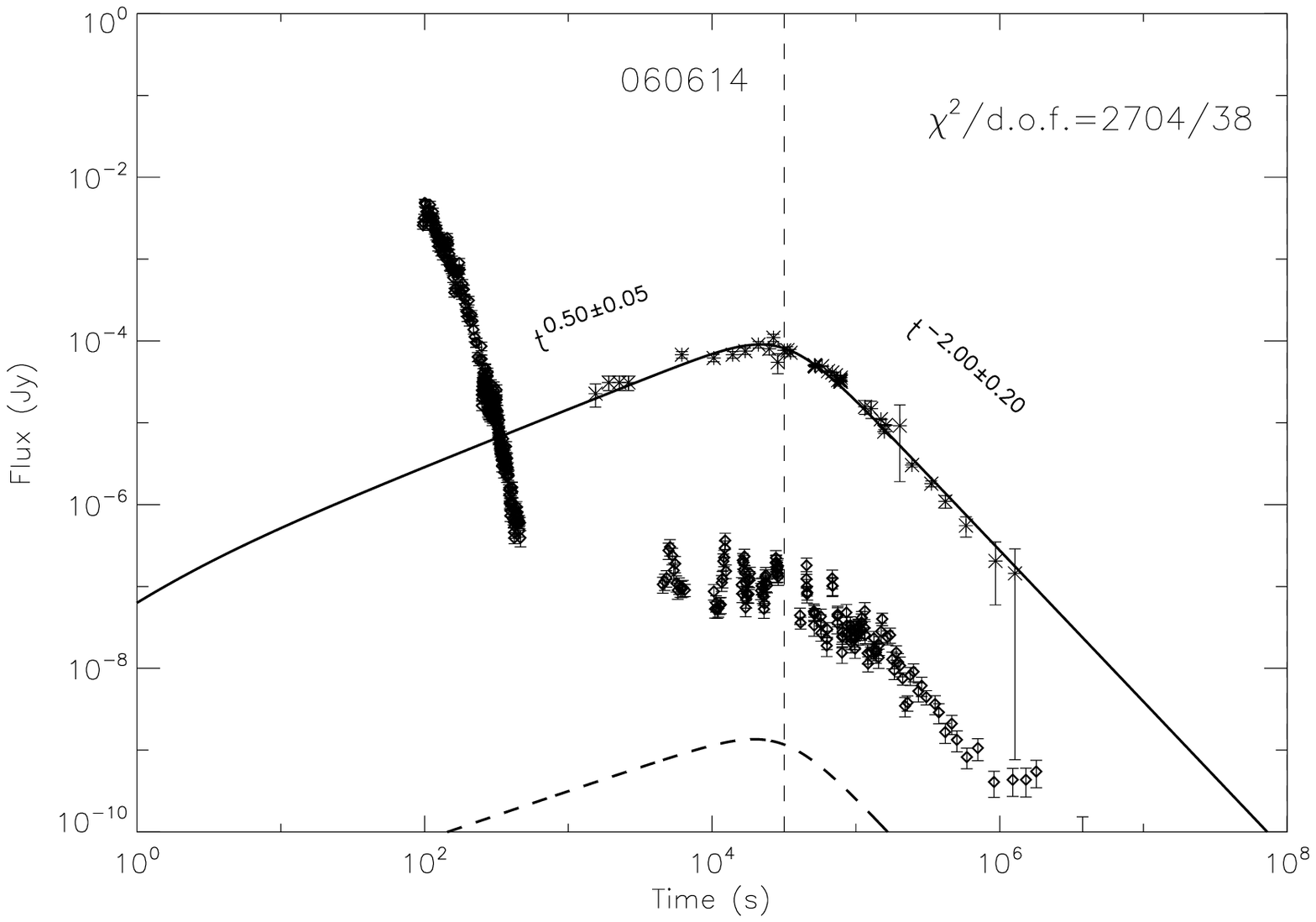}
\includegraphics[angle=0,scale=0.3]{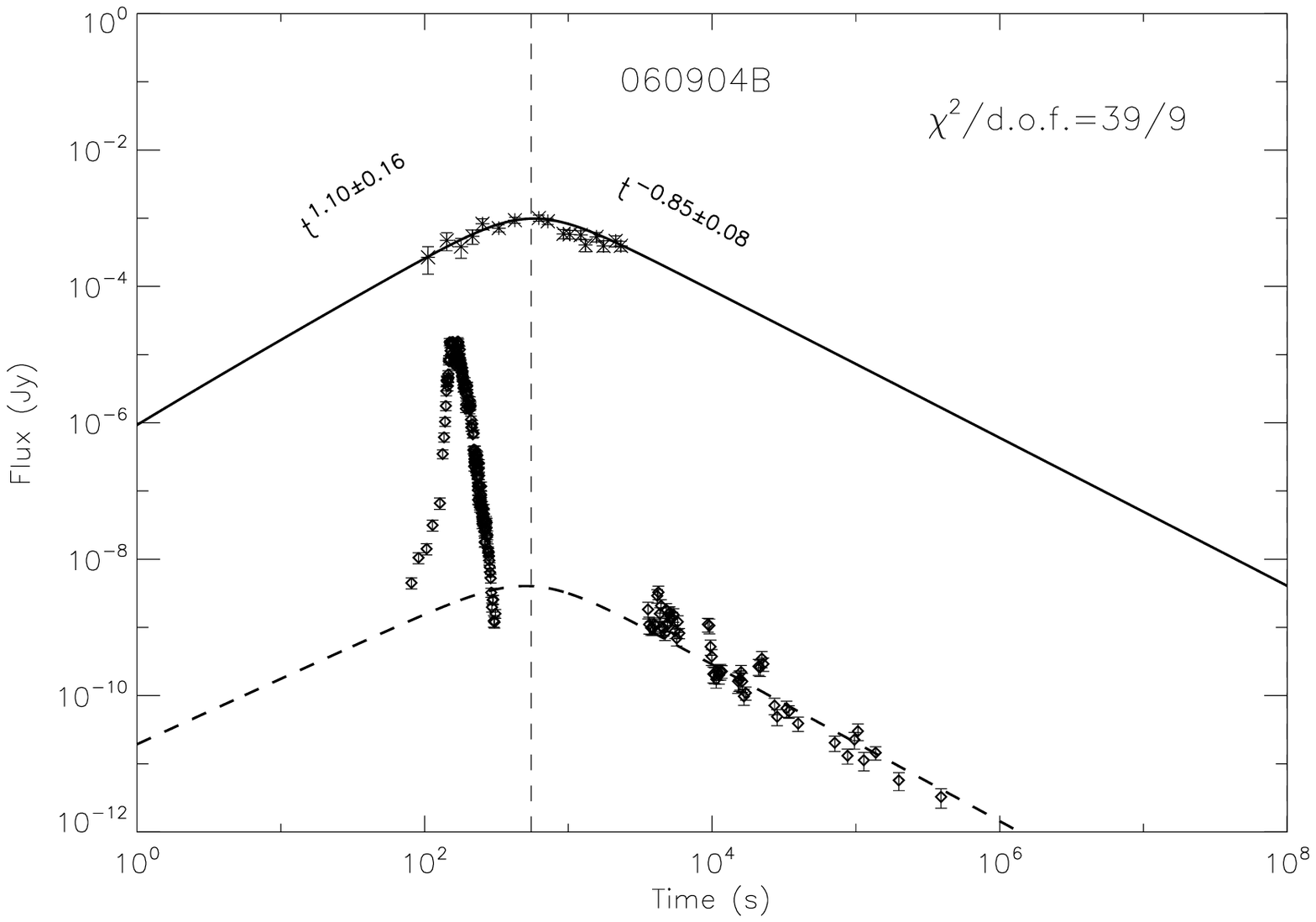}
\includegraphics[angle=0,scale=0.3]{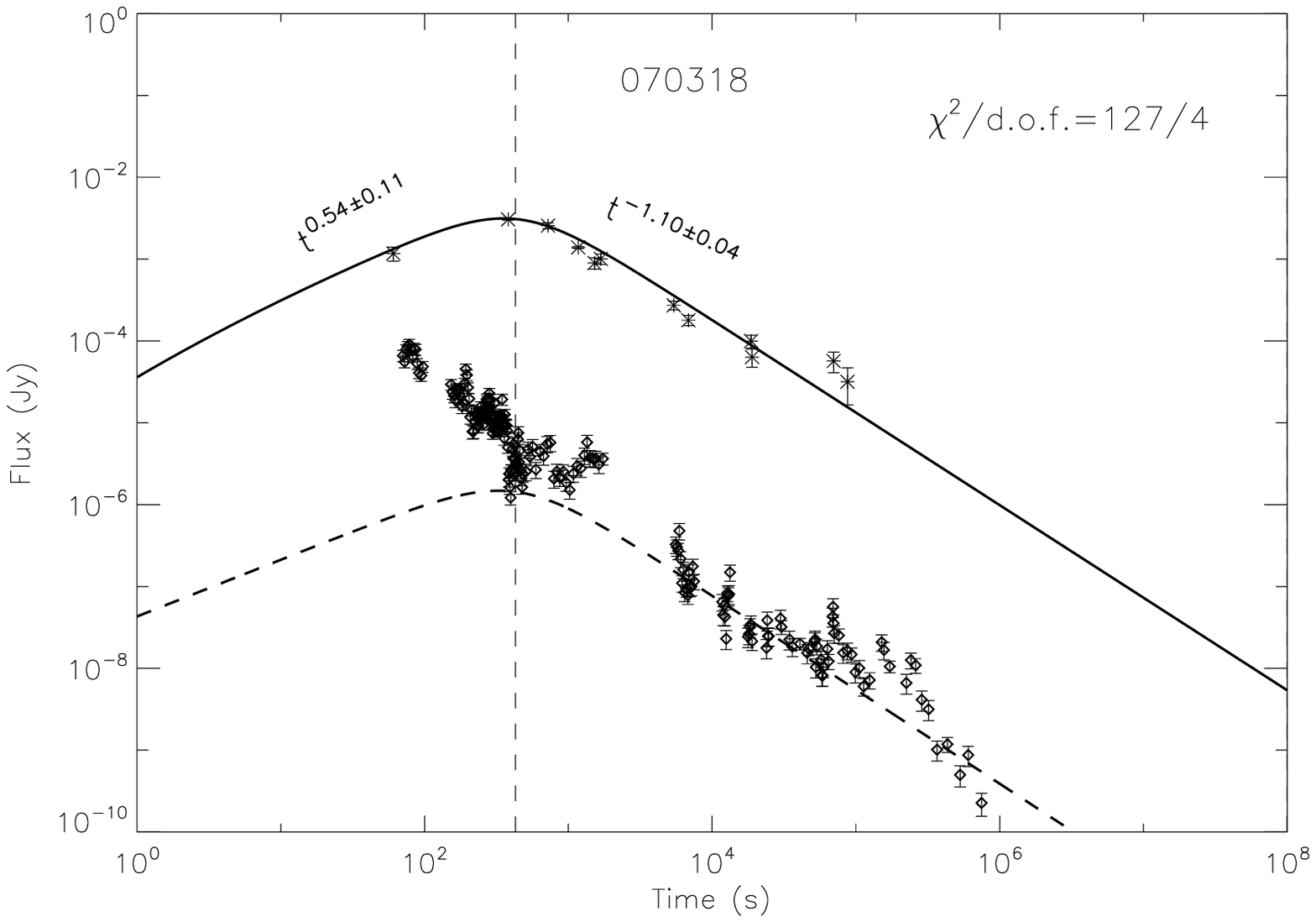}
\includegraphics[angle=0,scale=0.3]{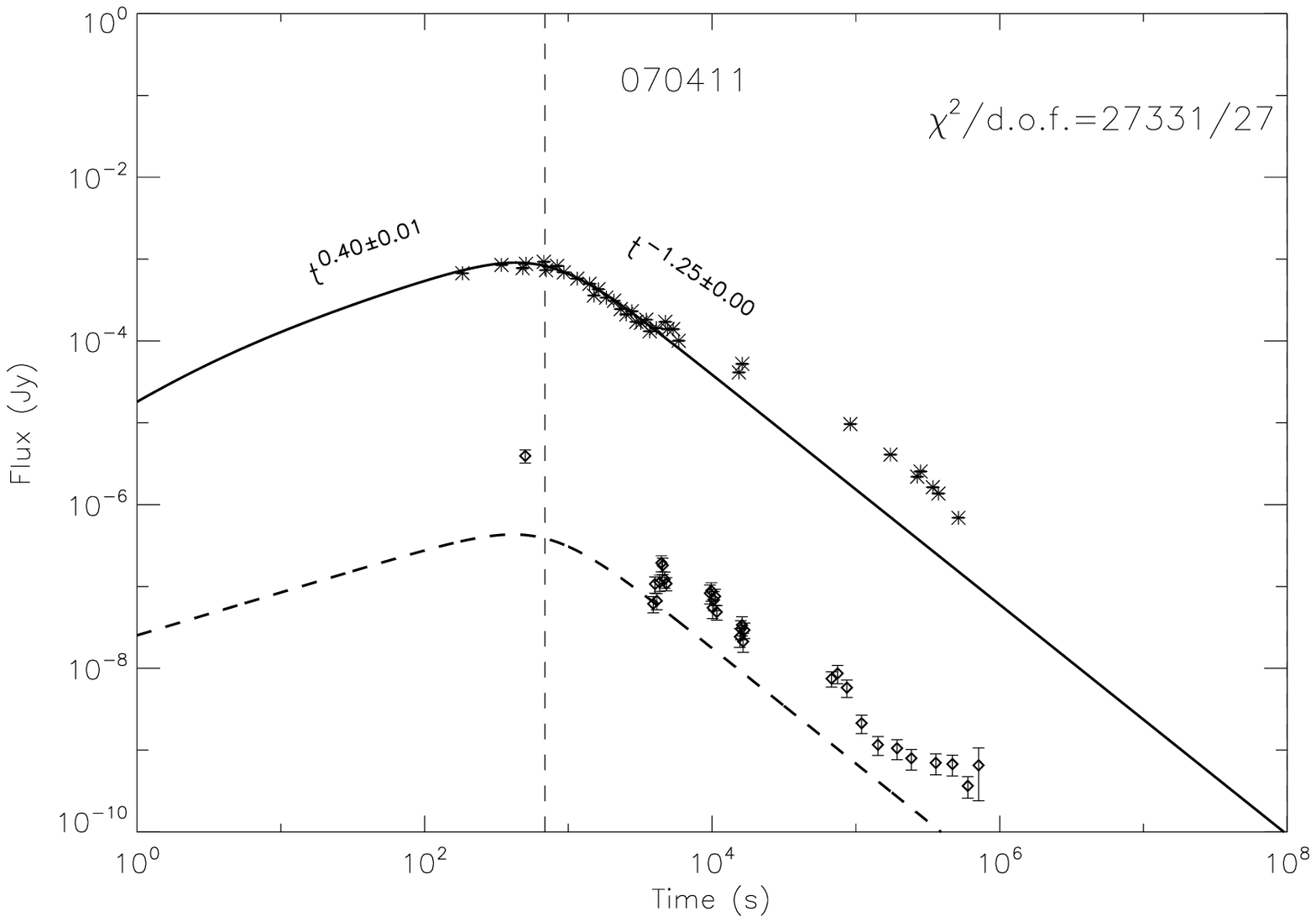}
\includegraphics[angle=0,scale=0.3]{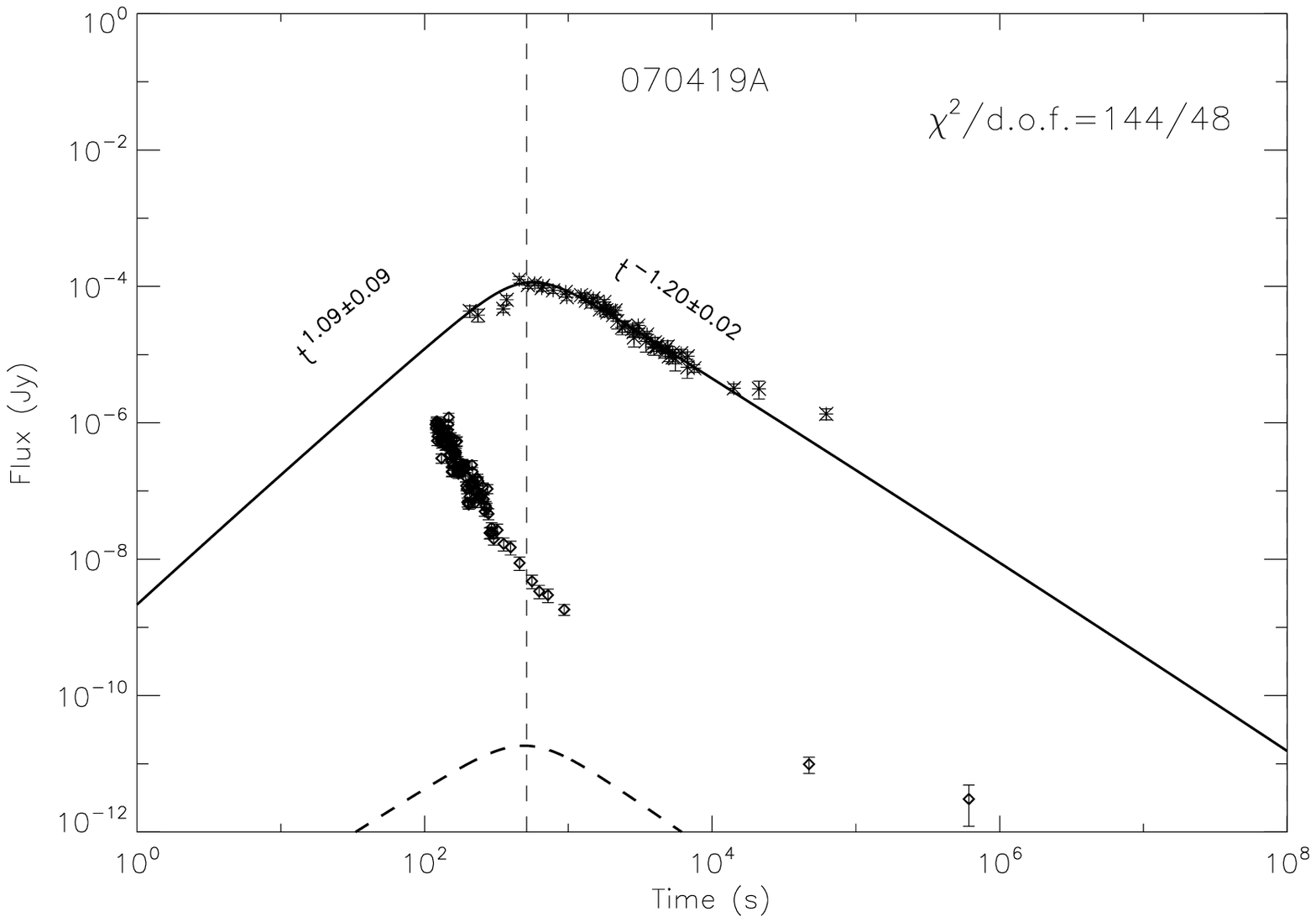}
\includegraphics[angle=0,scale=0.3]{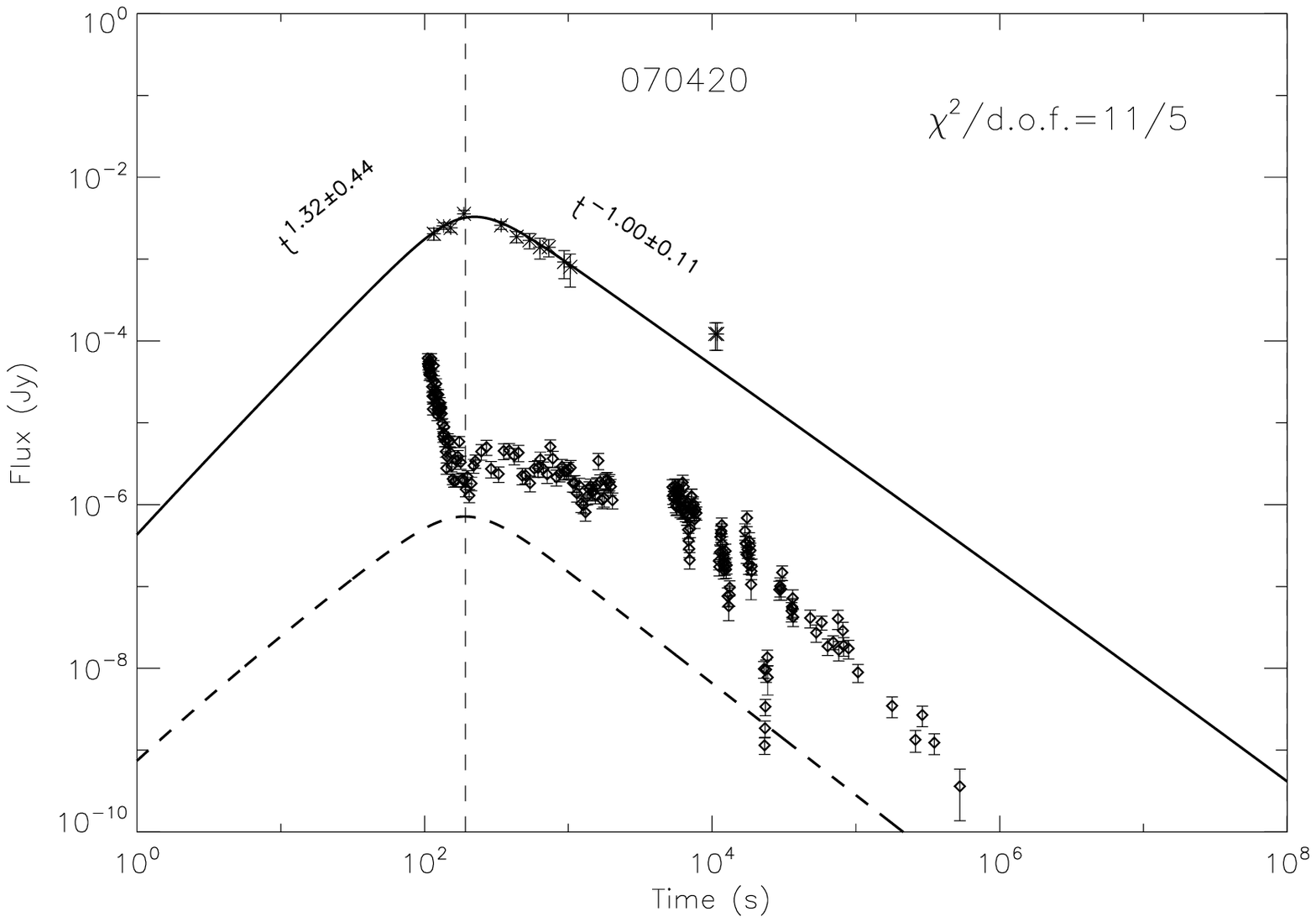}\hfill
\includegraphics[angle=0,scale=0.3]{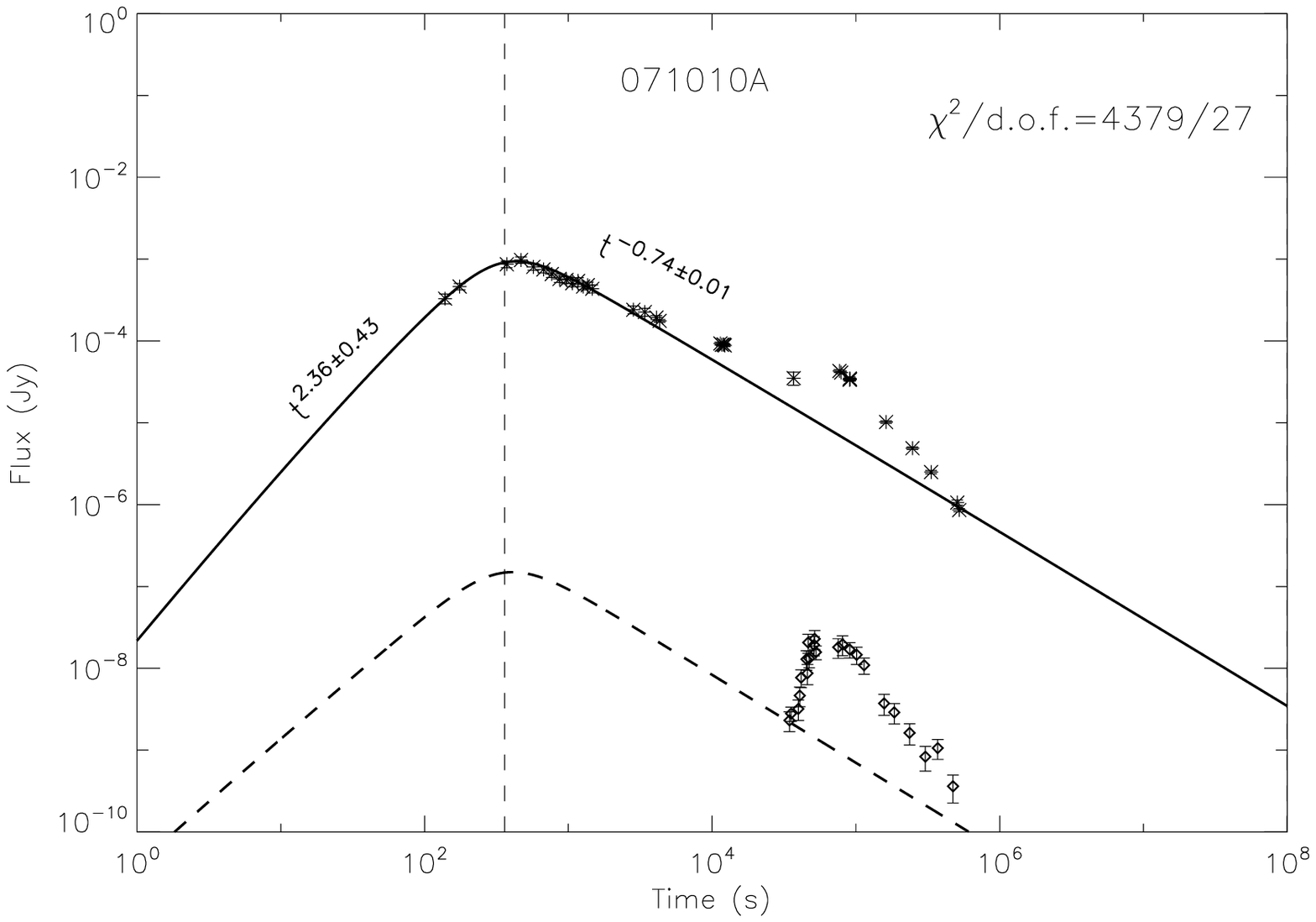}
\caption{The remaining GRBs of our selected sample. The symbols and
line styles are the same as Figure 3. The observed X-ray fluxes for
GRBs 060904B, 070419A, and 100906A, shown in this figure, have been
divided by 100.}
\end{figure*}

\begin{figure*}
\includegraphics[angle=0,scale=0.3]{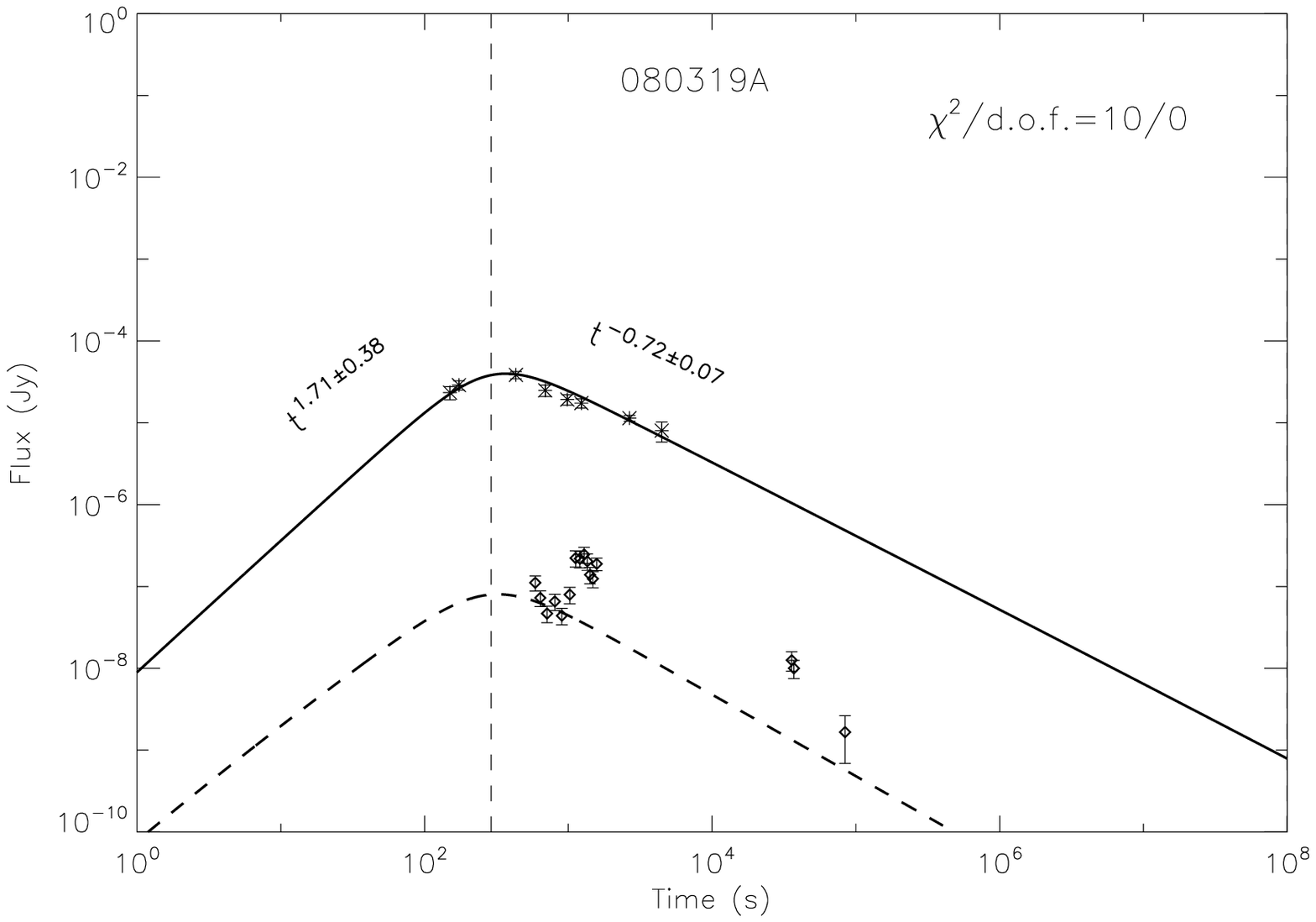}
\includegraphics[angle=0,scale=0.3]{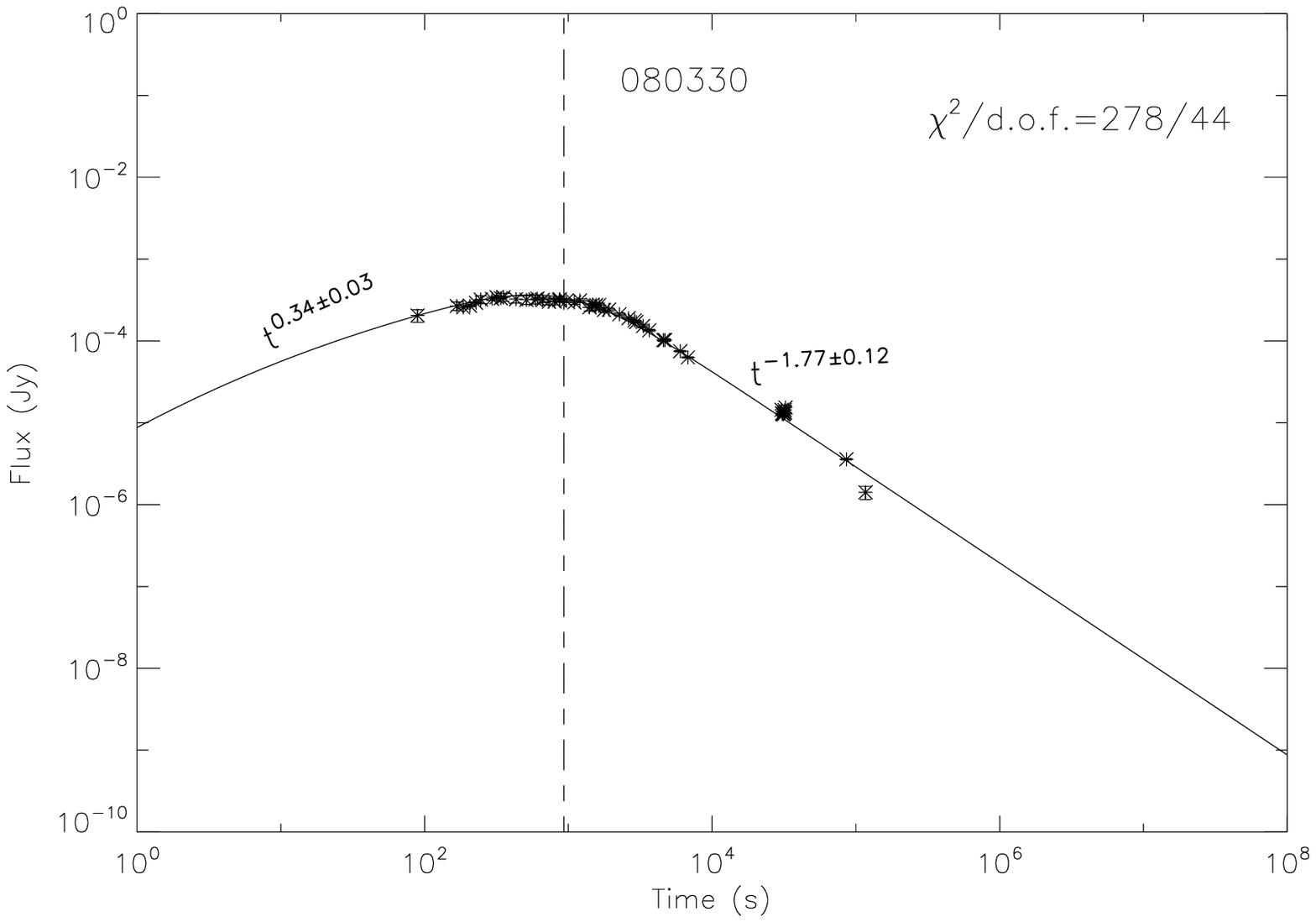}
\includegraphics[angle=0,scale=0.3]{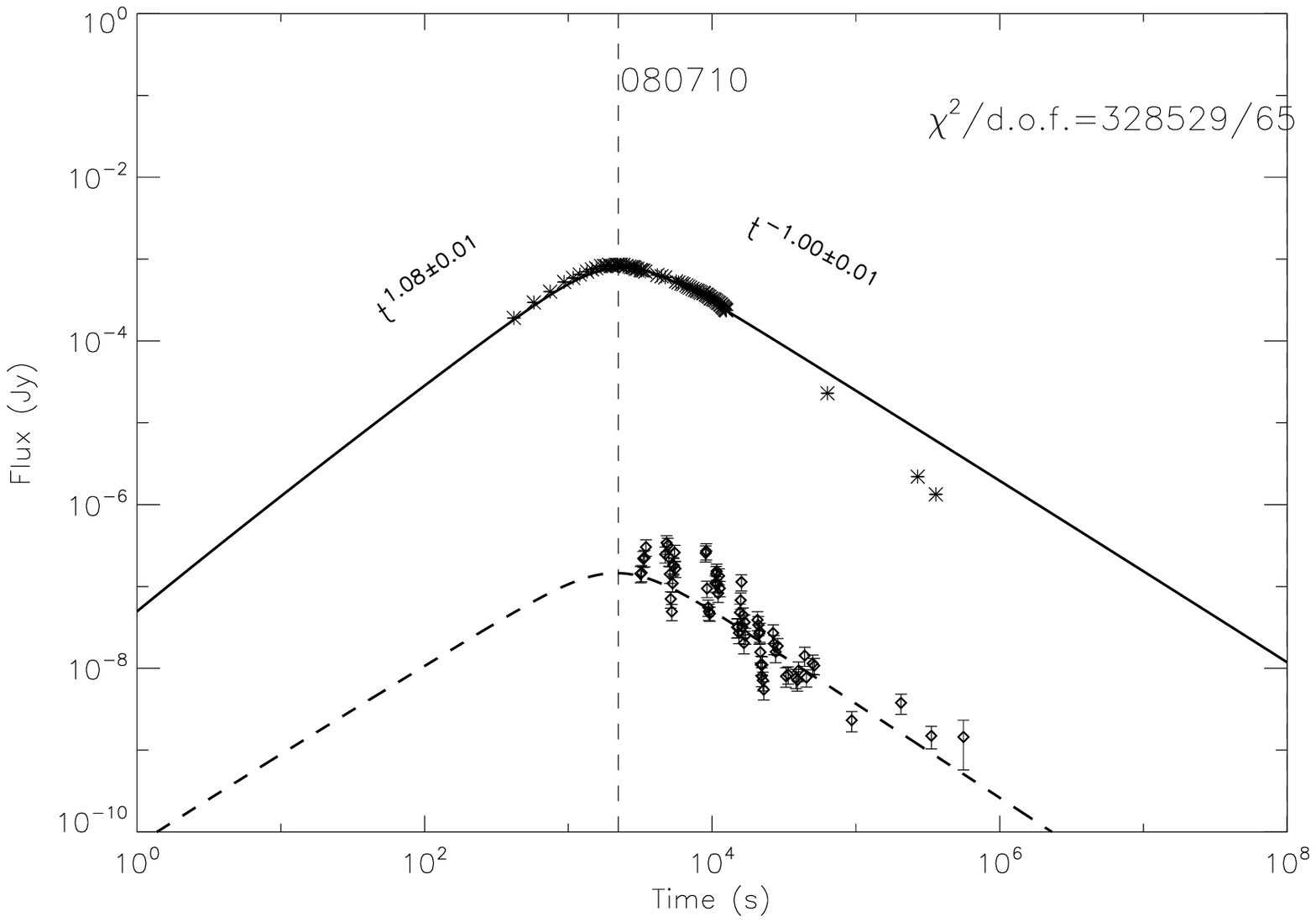}
\includegraphics[angle=0,scale=0.3]{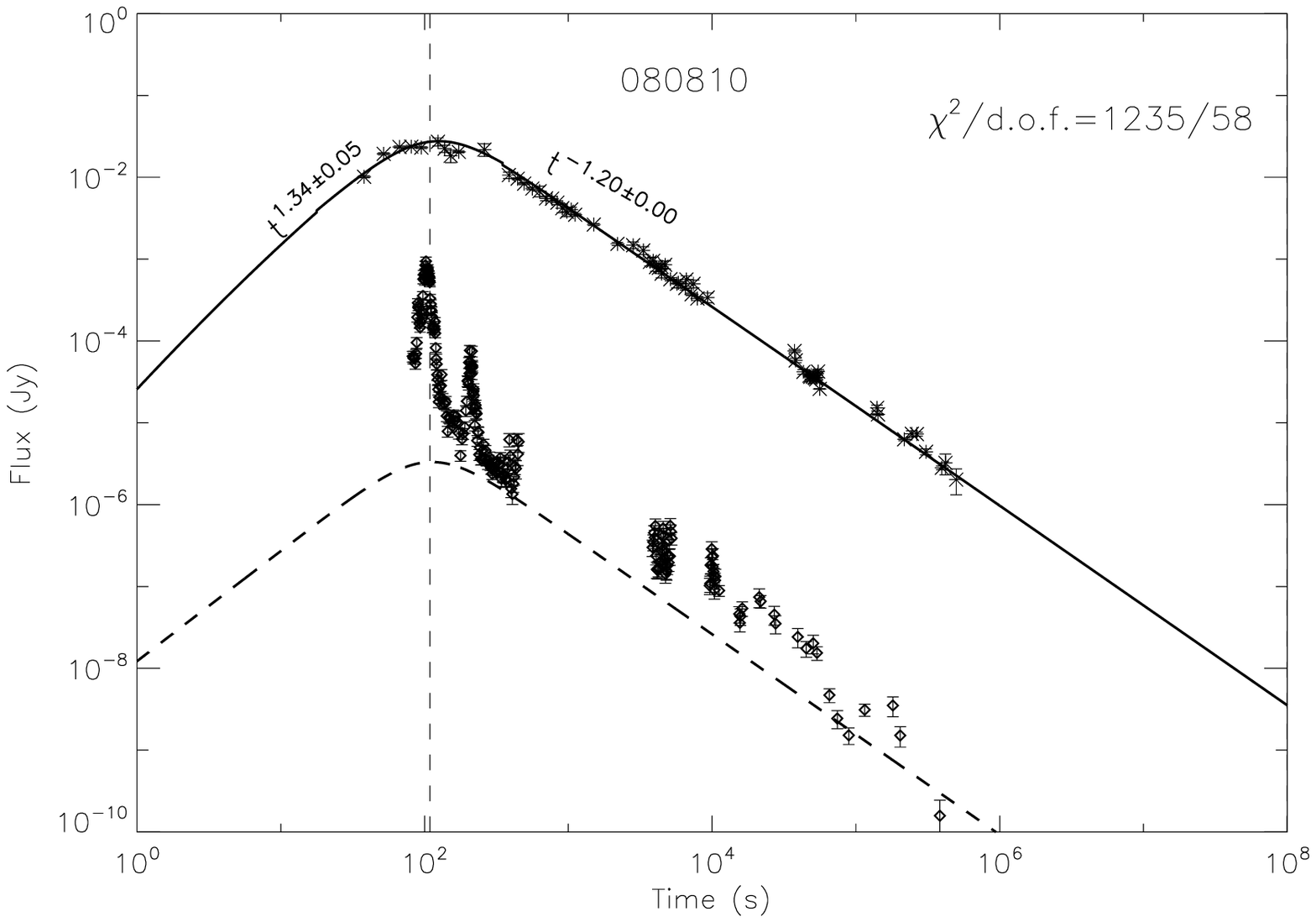}
\includegraphics[angle=0,scale=0.3]{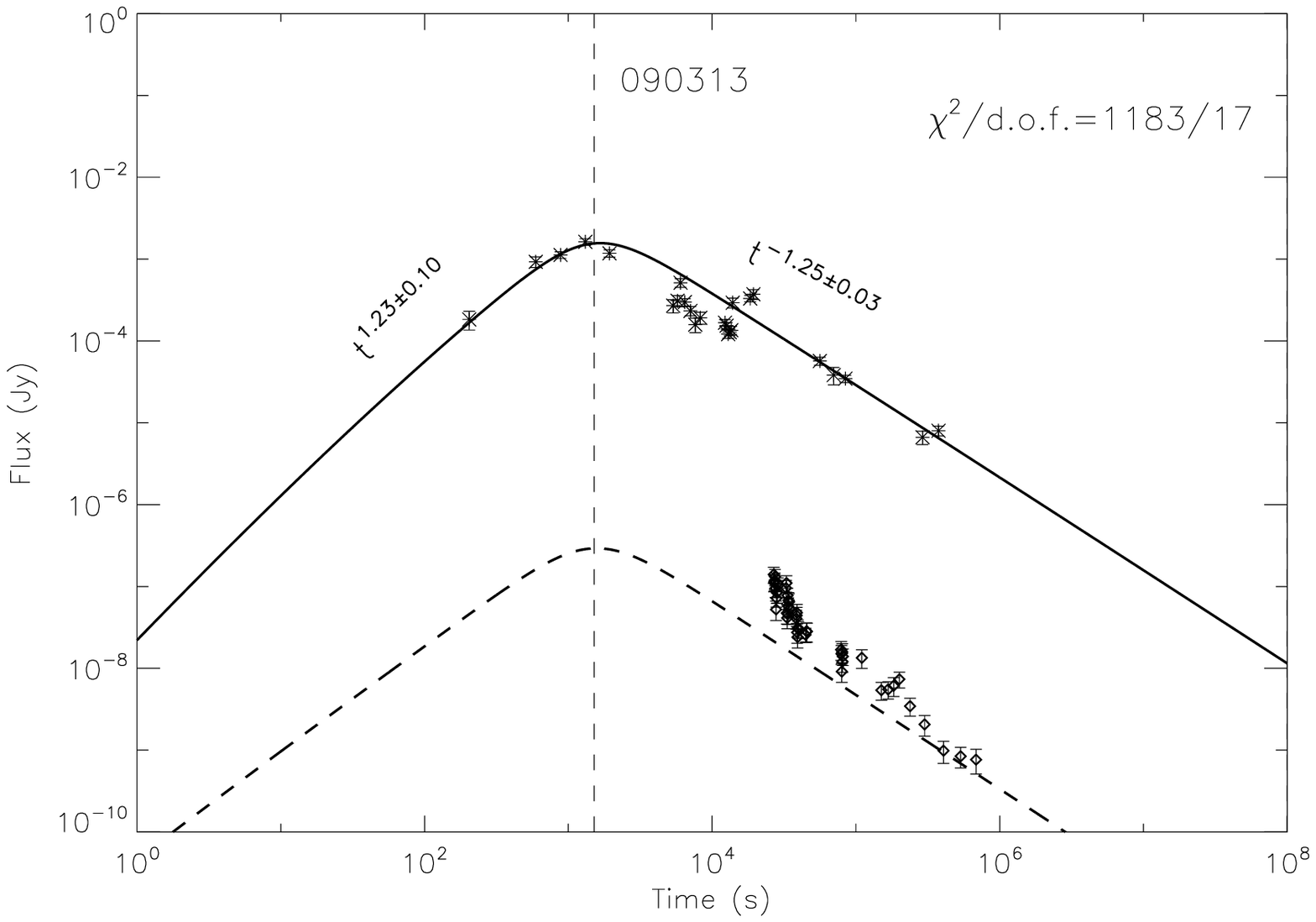}
\includegraphics[angle=0,scale=0.3]{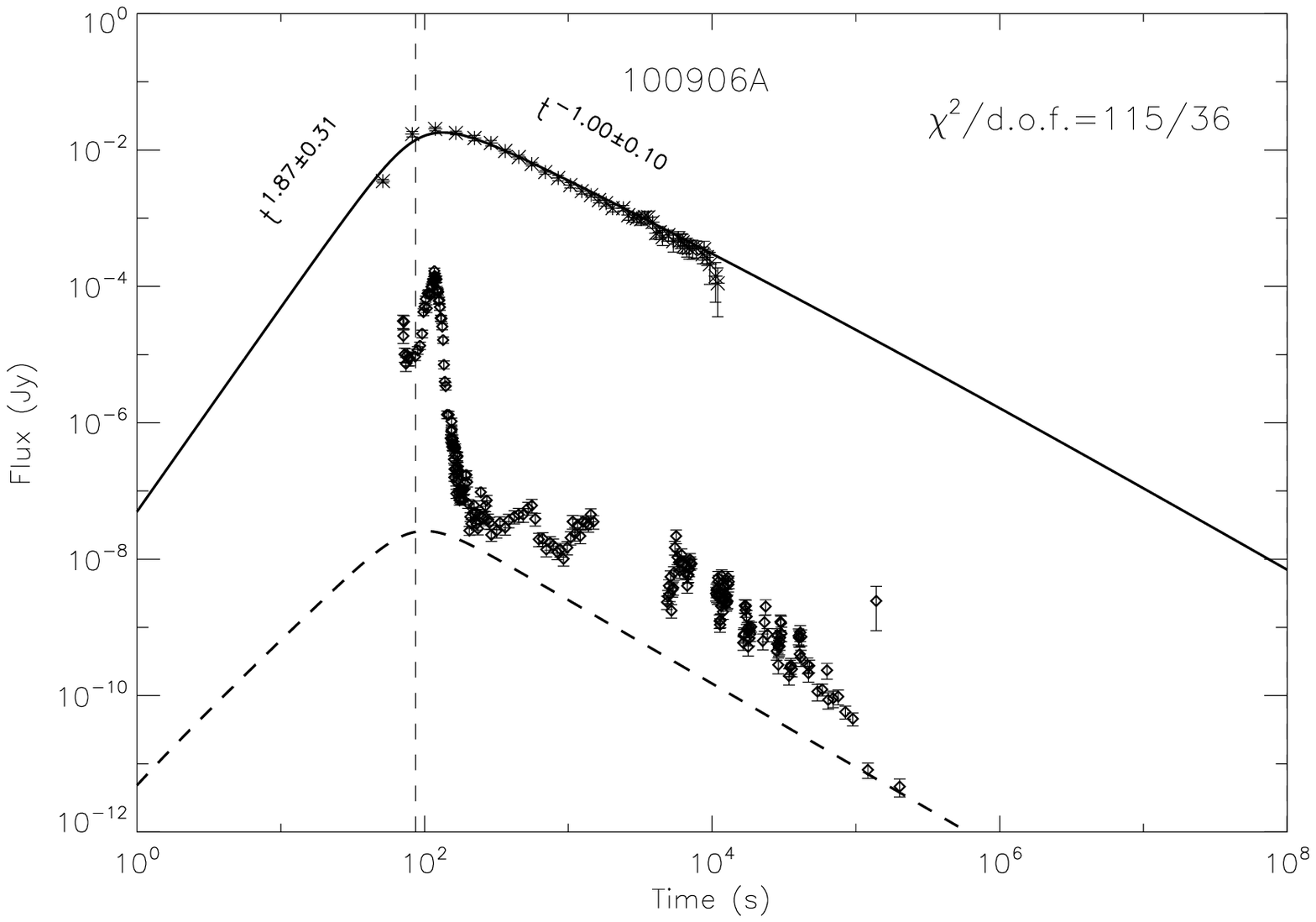}
\includegraphics[angle=0,scale=0.3]{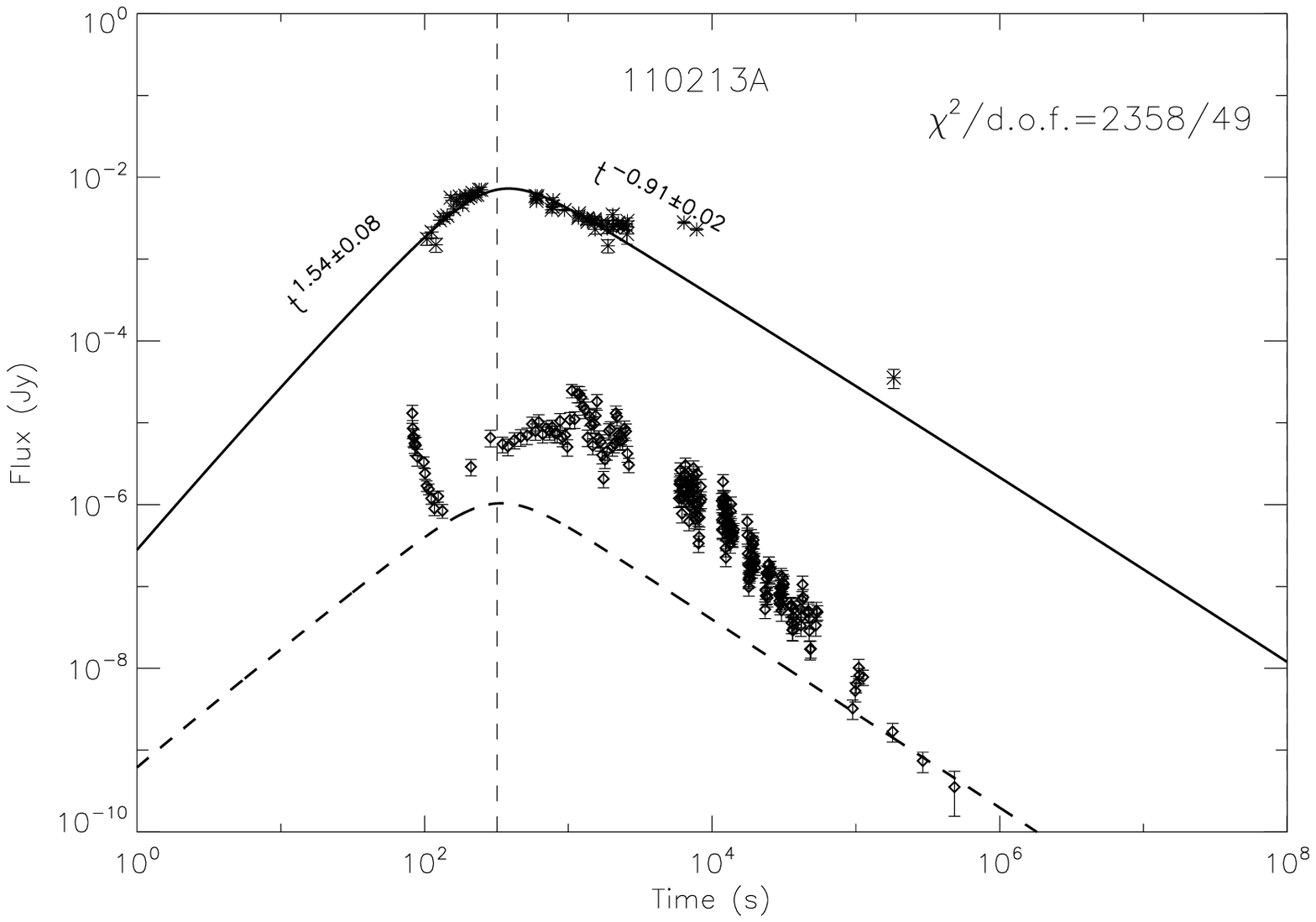}
\center{Fig. 4--- Continued}
\end{figure*}

\clearpage
\begin{figure*}
\includegraphics[angle=0,scale=0.80]{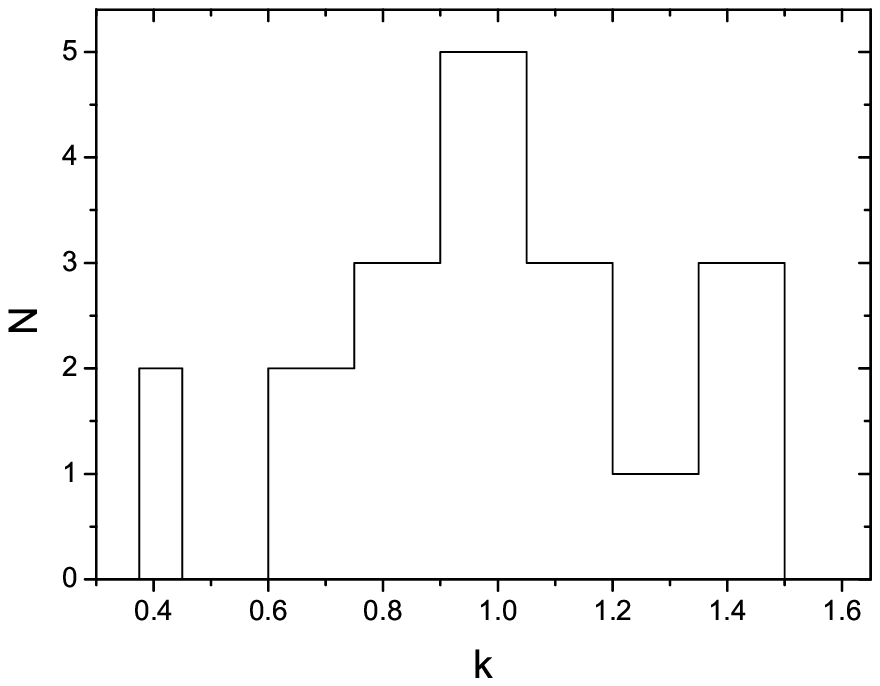}
\includegraphics[angle=0,scale=0.80]{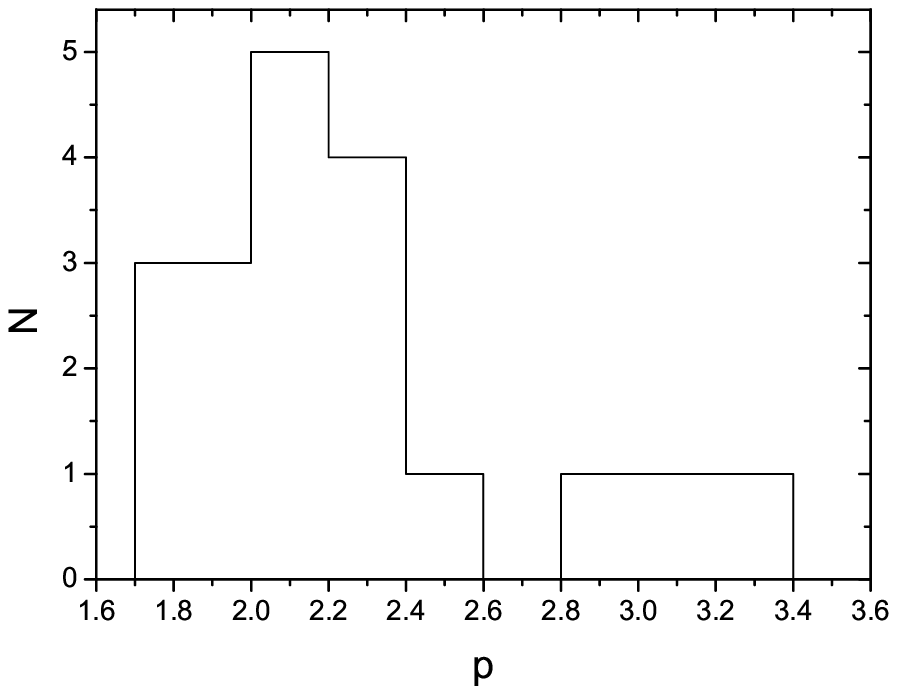}
\caption{Distributions of the values $k$ and $p$ of our sample. }
\end{figure*}

\clearpage
\begin{deluxetable}{lcccccccccccc}

\tablewidth{500pt} \tabletypesize{\tiny}
\tablecaption{Optical observations and fitting results for our
sample} \tablenum{1}

\tablehead{ \colhead{GRB}
&\colhead{$z$}
&\colhead{$\beta_{o}$}
&\colhead{$k$}
&\colhead{$p$}
&\colhead{$\varepsilon _{B}$}
&\colhead{$\varepsilon _{e}$}
&\colhead{$R_{0}$ (cm)}
&\colhead{$E$ (erg)}
&\colhead{$\eta$}
&\colhead{Emission regime}
&\colhead{Refs.}}

\startdata
030418  &   ...     &   ...                     &   1.09    $\pm$   0.12    &   1.73    $\pm$   0.11    &   0.2     &   0.2 &   2E16    &   2E52    &    75     &   $\nu _m^f < \nu < \nu_c^f $         &   (1) &\\
050730  &   3.969   &   0.56    $\pm$   0.06    &   0.92    $\pm$   0.11    &   2.16    $\pm$   0.23    &   0.1 &   0.15    &   3E17    &   4E53    &    105        &   $\nu _m^f < \nu < \nu_c^f $         &   (2) &\\
060605  &   3.773   &   1.04    $\pm$   0.05    &   1.08    $\pm$   0.11    &   2.08    $\pm$   0.10    &   0.2 &   0.02    &   1E17    &   8E53    &    120        &   $\nu  >\max\left\{ {\nu _c^f,\;\nu_m^f} \right\}$       &   (3, 4)  &\\
060614  &   0.125   &   0.94    $\pm$   0.08    &   1.19    $\pm$   0.05    &   3.38    $\pm$   0.28    &   0.2 &   0.02    &   1E15    &   1E53    &    30     &   $\nu _m^f < \nu < \nu_c^f $         &   (5) &\\
060904B &   0.703   &   1.11    $\pm$   0.1     &   0.95    $\pm$   0.17    &   1.80    $\pm$   0.11    &   0.01&   0.01    &   5E16    &   3E53    &    70     &   $\nu  >\max\left\{ {\nu _c^f,\;\nu_m^f} \right\}$       &   (6, 7)  &\\
070318  &   0.836   &   0.78                    &   1.38    $\pm$   0.06    &   2.11    $\pm$   0.06    &   0.01&   0.01    &   9E16    &   6E53    &    80     &   $\nu _m^f < \nu < \nu_c^f $         &   (8, 9)  &\\
070411  &   2.954   &   ...                     &   1.43    $\pm$   0.01    &   2.30    $\pm$   0.00    &   0.01    &   0.01    &   1E17    &   2E54     &  110     &   $\nu _m^f < \nu < \nu_c^f $         &   (10)    &\\
070419A &   0.97    &   0.82    $\pm$   0.16    &   1.04    $\pm$   0.05    &   2.37    $\pm$   0.03    &   0.1 &   0.01    &   4E15    &   1E52    &    60     &   $\nu _m^f < \nu < \nu_c^f $         &   (11, 12)    &\\
070420  &   ...     &   ...                     &   0.94    $\pm$   0.25    &   2.13    $\pm$   0.17    &   0.01&   0.01    &   5E16    &   6E52    &    85     &   $\nu _m^f < \nu < \nu_c^f $         &   (6) &\\
071010A &   0.98    &   0.76    $\pm$   0.23    &   0.37    $\pm$   0.25    &   1.92    $\pm$   0.05    &   0.3 &   0.01    &   3E16    &   6E52    &    70     &   $\nu _m^f < \nu < \nu_c^f $         &   (13, 12)    &\\
071031  &   2.05    &   0.9     $\pm$   0.1     &   1.40     $\pm$  0.00    &   1.79    $\pm$   0.00        &   0.2 &   0.02    &   2E16    &   5E53     &  90      &   $\nu  >\max\left\{ {\nu _c^f,\;\nu_m^f} \right\}$       &   (14)    &\\
080319A &   ...     &   0.77    $\pm$   0.02    &   0.76    $\pm$   0.22    &   1.80    $\pm$   0.11    &   0.1 &   0.1     &   1E15    &   5E51    &    80     &   $\nu _m^f < \nu < \nu_c^f $         &   (15)    &\\
080330  &   1.51    &   0.99                    &   1.32    $\pm$   0.03    &   3.03    $\pm$   0.16    &   0.02&   0.02    &   4E16    &   4E53    &    80     &   $\nu _m^f < \nu < \nu_c^f $         &   (16, 9) &\\
080710  &   0.845   &   1.00    $\pm$   0.02    &   0.92    $\pm$   0.00    &   2.00    $\pm$   0.00        &   0.2 &   0.01    &   1E16    &   4E53     &  60      &   $\nu  >\max\left\{ {\nu _c^f,\;\nu_m^f} \right\}$       &   (17, 12)    &\\
080810  &   3.35    &   0.51    $\pm$   0.22    &   0.90    $\pm$   0.03    &   2.41    $\pm$   0.01    &   0.05&   0.04    &   5E17    &   4E54    &    170        &   $\nu _m^f < \nu < \nu_c^f $         &   (18)    &\\
081203A &   2.05    &   0.9     $\pm$   0.01    &   0.40    $\pm$   0.01    &   2.91    $\pm$   0.01    &   0.01&   0.01    &   1E17    &   2E54    &    120        &   $\nu _m^f < \nu < \nu_c^f $         &   (19)    &\\
090313  &   3.375   &   1.2                 &   0.71    $\pm$   0.09    &   2.33    $\pm$   0.04    &   0.1 &   0.01    &   8E16    &   3E54    &    90     &   $\nu  >\max\left\{ {\nu _c^f,\;\nu_m^f} \right\}$       &   (20)    &\\
100906A &   1.727   &   ...                     &   0.63    $\pm$   0.17    &   2.21    $\pm$   0.14    &   0.01&   0.01    &   1E17    &   8E53    &    180        &   $\nu _m^f < \nu < \nu_c^f $         &   (22)    &\\
110213A &   1.46    &   ...                     &   0.83    $\pm$   0.04    &   2.04    $\pm$   0.03    &   0.1 &   0.01    &   4E16    &   6E53    &    110        &   $\nu _m^f < \nu < \nu_c^f $         &   (23)    &\\
\enddata

\tablerefs{(1) Rykoff et al. 2004; (2) Pandey et al. 2006; (3)
Rykoff et al. 2009; (4) Ferrero et al. 2009; (5) Della Valle et al.
2006; (6) Klotz et al. 2008; (7) Kann et al. 2010; (8) Roming et al.
2009; (9) Fynbo et al. 2009; (10) Ferrero et al. 2008; (11) Melandri
et al. 2009; (12) Liang et al. 2010; (13) Covino et al. 2008; (14)
Kr{\"u}hler et al. 2009a; (15) Li et al. 2012 (16) Guidorzi et al.
2009; (17) Kr{\"u}hler et al. 2009b; (18) Page et al. 2009; (19)
Kuin et al. 2009; (20) Melandri et al. 2010; (21) Gorbovskoy et al.
2012; (22) Gorbovskoy et al. 2012; (23) Liang et al. 2013 }

\end{deluxetable}

\end{document}